\documentclass{article}
\PassOptionsToPackage{numbers, compress}{natbib}
\usepackage[main, final]{neurips_2026}

\usepackage{float}
\usepackage[utf8]{inputenc} 
\usepackage[T1]{fontenc}    
\usepackage{hyperref}       
\usepackage{url}            
\usepackage{booktabs}       
\usepackage{amsfonts}       
\usepackage{amsmath}
\usepackage{amssymb}
\usepackage{nicefrac}       
\usepackage{microtype}      
\usepackage{xcolor}         
\usepackage{xspace}
\usepackage{array}
\usepackage{tabularx}
\usepackage{makecell}
\usepackage{multirow}
\usepackage{algorithm}
\usepackage{algorithmic}
\usepackage{tikz}
\usetikzlibrary{positioning,arrows.meta,calc,fit,backgrounds,shapes.geometric,decorations.pathreplacing,matrix,patterns}
\usepackage{graphicx}
\usepackage{wrapfig}
\usepackage{enumitem}
\usepackage[labelformat=simple,labelsep=space]{subcaption}

\usepackage[capitalize,noabbrev]{cleveref}
\usepackage{tcolorbox}

\newcommand{\framework}{\textsc{ReCoVer}\xspace}
\newcommand{\ulfmallreduce}{\textsc{ulfm\_allreduce}\xspace}
\newcommand{\ulfmconsensus}{\textsc{ulfm\_consensus}\xspace}
\newcommand{\refalg}[2]{\hyperref[#1]{#2}\xspace}
\newcommand{\callulfmallreduce}{\refalg{alg:ulfm-allreduce}{\textsc{ulfm\_allreduce}}}
\newcommand{\callulfmconsensus}{\refalg{alg:ulfm-consensus}{\textsc{ulfm\_consensus}}}
\newcommand{\callhandlefail}{\refalg{alg:handle-failure}{\textsc{handle\_work\_failure}}}
\newcommand{\callgradrestore}{\refalg{alg:gradient-restoration}{\textsc{gradient\_restoration}}}
\newcommand{\callpolicyadjust}{\refalg{alg:policy-adjustment}{\textsc{policy\_adjustment}}}
\newcommand{\callpolicyadvance}{\refalg{alg:policy-advancement}{\textsc{policy\_advancement}}}

\title{\framework: Resilient LLM Pre-Training System via Fault-Tolerant Collective and Versatile Workload}

\author{
  {\bf Ziyue Liu}\textsuperscript{1}, \ {\bf Zhengyang Wang}\textsuperscript{1}, \ {\bf Ruijie Zhang}\textsuperscript{1}, \ {\bf Avinash Maurya}\textsuperscript{2}, \ {\bf Hui Zhou}\textsuperscript{2}, \\ {\bf Paul Hovland\textsuperscript{2}}, \ {\bf Sheng Di\textsuperscript{2}}, \ {\bf Franck Cappello\textsuperscript{2}}, \ {\bf Bogdan Nicolae\textsuperscript{2}}, \ {\bf Zheng Zhang\textsuperscript{1}} \\
  \textsuperscript{1}University of California at Santa Barbara; \textsuperscript{2}Argonne National Laboratory\\
  \texttt{ziyueliu@ucsb.edu}, \ zhengzhang@ece.ucsb.edu \\
}

\begin{document}

\maketitle

\begin{abstract}
Pre-training large language models on massive GPU clusters has made hardware faults routine rather than rare, driving the need for resilient training systems. Yet existing frameworks either focus on specific parallelism schemes or risk drifting away from a failure-free training trajectory. We propose \framework, a resilient LLM pre-training system that upholds a single invariant: each iteration keeps the number of microbatches constant, ensuring per-iteration gradients remain stochastically equivalent to a failure-free run. The framework is organized as three decoupled protocol layers: (1) Fault-tolerant collectives that isolate faults from propagating across replicas; (2) in-step fine-grained recovery that preserves intra-iteration progress and prevents gradient corruption; (3) versatile-workload policy that dynamically redistributes microbatch quotas across the survivors. The design is parallelism-agnostic, integrating directly with both 3D parallelism and Hybrid Sharded Data Parallel (HSDP) as a drop-in substrate. We evaluate our implementation on end-to-end pre-training tasks for up to 512 GPUs, \framework successfully preserves the training trajectory from a failure-free reference despite 256 GPUs lost spread across the run. For comparison with checkpoint-and-restart baselines, \framework demonstrates $2.23\times$  higher effective throughput after successive failures. This advantage results in \framework processing $74.9\%$ more tokens at 234 GPU-hours, with the gap widening as the training prolongs~\footnote{Code is available at \href{https://github.com/alvin-zyl/ReCoVer}{https://github.com/alvin-zyl/ReCoVer}.}.
\end{abstract}

\section{Introduction}
\label{sec:intro}

The increasing demand of pre-training frontier large language models (LLMs) has driven industrial high-performance computing (HPC) platforms to rapidly scale to over $O(100\text{k})$ GPUs. For instance, LLaMA-4 was trained on 100k GPUs~\cite{llama4100k}; Grok-4 was trained on 200k GPUs~\cite{grok4}, while xAI was reported to build a cluster with 1 million H100 GPUs~\cite{1mgpuxai}. At such scales, the mean-time-between-failures (MTBF) shrinks inversely with the GPU count: from 3 hours at 16k GPUs~\cite{llama3} to only 18 min at 100k GPUs~\cite{salpekar2026training}, and projected below 5 min at 600k~\cite{lee2026spare}. Thus,
traditional checkpoint-restart techniques are bottlenecked by \emph{recovery overhead} (re-initialize the communication backend and training pipeline, load model parameters and optimizer state, replay from last checkpoint), which is reported to be 10 min on production 100k-GPU clusters where failure lands every 18 min~\cite{salpekar2026training}. That is, system makes progress in only 8 of every 18 min, $>50\%$ of total GPU hours are wasted. Despite significant progress in reducing checkpoint overhead~\cite{mohan2021checkfreq,eisenman2022check,wang2023gemini,maurya2024datastates,nicolae2019veloc,wan2024bytecheckpoint,wan2025robust,maurya2026datastatesllm}, we still approach zero useful throughput when MTBF approaches the restart overhead, motivating the need for alternatives that keep the job alive across failures, which we call \emph{forward recovery} methods.


{\bf Limitations of state of the art.}
Although forward recovery has been studied in the AI and HPC community, existing efforts fall short along three dimensions. First, prior work is often \emph{\color{blue}layer-isolated}: resilience
of the communication layer~\cite{ulfm,ulfm-mpich,bland2013post,bouteiller2015plan,laguna2014evaluating,li2023elastic} is well studied but insufficient in today's highly structured LLM pre-training stacks~\cite{shoeybi2019megatron,narayanan2021efficient,huang2019gpipe,rajbhandari2020zero,zhao2023pytorch,jiang2024megascale} where communication patterns involve more complex multi-collective, multi-process-group with
strict performance guarantees. Second, prior work lacks \emph{\color{blue}versatility}: pipeline-centric systems~\cite{thorpe2023bamboo,jang2023oobleck,gandhi2024recycle} embed shadow stages into the pipeline schedule and are tightly coupled to that single parallelism flavor, failing to extend to other parallelism schemes, such as HSDP~\cite{zhao2023pytorch}. Third, in a quest to lower performance overheads,
some approaches do not preserve \emph{\color{blue}computational equivalence}: \cite{salpekar2026training} continues training with reduced number of microbatches after a failure, which shifts the gradient-noise scale and induces loss spikes that, under frequent failures, may accumulate and drift away from the failure-free baseline.
A recent proposal~\cite{lee2026spare} pursues forward recovery via data shard replication, but its performance is only validated via simulation without providing a real implementation, and the method undergoes a large factor of redundant computation once a failure occurs.

{\bf Contributions.}
We present \framework, a resilient LLM pre-training system that closes the three gaps above. \framework delivers \emph{forward recovery} that (i)~is \emph{extended} across the full pre-training stack rather than the communication layer alone, (ii)~is \emph{versatile} across different parallelism schemes, and (iii)~preserves \emph{computational equivalence} to the failure-free trajectory, regardless of when or where a failure occurs. This is realized as a three-layer fault-tolerant protocol that, under any failure schedule, upholds a single invariant: every iteration commits gradients from the same number of microbatches as its failure-free reference, keeping each optimizer update stochastically equivalent, without rolling back, replaying, or paying for idle replication. We summarize our contributions as follows:

\begin{itemize}[leftmargin=*]
\item {\bf Analysis of fault-tolerant pre-training (\cref{sec:challenges}).} We examine a synchronous iteration under device loss, analyze how a failure propagates through collective communication, and identify five challenges that any forward-recovery system must address.
\item {\bf Bottom layer: contains failures locally (\cref{sec:ulfm-guarded}).} Going beyond communication-layer-only resilience, \framework extends fail-continue guarantees to the multi-collective, multi-process-group structure of modern LLM pre-training: it repairs communicators in-place over survivors and constructs a collectively agreed post-failure consistent view.
\item {\bf Middle layer: enables \emph{forward recovery} (\cref{sec:instep}).} \framework orchestrates training-level recovery within the failed iteration, restoring affected gradients to their pre-reduction state to prevent corruption and preserve survivor's intra-iteration progress, without rolling back or replaying.
\item {\bf Top layer: guarantees \emph{computational equivalence} (\cref{sec:versatile,app:proof}).} \framework dynamically redistributes microbatch quotas across survivors to keep the total number of microbatches constant, thus being stochastically equivalent to its failure-free trajectory without pre-allocated idle replicas.
\item {\bf Integration and evaluation across parallelism stacks (\cref{sec:integration,sec:eval}).}
Against checkpoint-restart baselines, \framework delivers $2.23\times$ higher effective throughput under successive failures and processes $74.9\%$ more tokens at 234 GPU-hours, with the gap widening as training prolongs, while preserving training trajectory to its failure-free reference despite losing up to 256 GPUs.
\end{itemize}

\section{Related Work}
\label{sec:related}

{\bf Failure issues in LLM pre-training systems.}
Modern LLM pre-training rests on a decade of parallel-training
infrastructure~\cite{sergeev2018horovod,shoeybi2019megatron,narayanan2021efficient,huang2019gpipe,narayanan2019pipedream,rajbhandari2020zero,zhao2023pytorch,jiang2024megascale,dash2024optimizing,wang2025boost}. All devices
rendezvous at gradient synchronization, so a single device's failure terminates the job. Production-cluster studies~\cite{jeon2019analysis,hu2024characterization,kokolis2025revisiting,xiong2024superbench,salpekar2026training} report mean-time-between-failures (MTBF) to be tens of minutes at the 100k-GPU scale. At such scale, effective training time is less than even 50\%~\cite{salpekar2026training}, making fault tolerance a first-class system requirement. This issue is expected to become much more significant as the cluster continues to scale up and becomes more heterogeneous.

{\bf Checkpoint and restart.}
The prevailing resilience strategy is periodic checkpointing of model, optimizer, and dataloader state combined with job restart from the latest checkpoint on failure~\cite{narayanan2021efficient,jiang2024megascale}. Subsequent systems lower checkpoint overhead through frequent fine-grained snapshots~\cite{mohan2021checkfreq}, in-memory and hierarchical replication~\cite{wang2023gemini,eisenman2022check}, lazy asynchronous
persistence~\cite{maurya2024datastates,nicolae2019veloc,maurya2026datastatesllm}, and unified multi-framework checkpoint management~\cite{wan2024bytecheckpoint}. All retain the restart-and-replay semantics.

{\bf Fault-tolerant frameworks.}
A second line of research keeps the job alive across failures. At the
communication layer, User-Level Failure Mitigation (ULFM)~\cite{ulfm,ulfm-mpich,bland2013post,bouteiller2015plan,laguna2014evaluating}
lifts MPI from fail-stop to fail-continue; Early work~\cite{li2023elastic} integrates it with Horovod~\cite{sergeev2018horovod} for DP-only deep learning applications. R$^2$CCL~\cite{wang2025reliable} handles NCCL link failures, not process crashes. Pipeline-centric
systems~\cite{thorpe2023bamboo,jang2023oobleck,gandhi2024recycle} make pipeline stages the unit of redundancy, running shadow stages or rebuilding the schedule around survivors. In production, FTAR~\cite{salpekar2026training} extends NCCL with a revoke-and-rejoin path and reports HSDP pre-training through sustained failures by re-provisioning a replica and asynchronously rejoining it.

\section{Fault-Tolerant Challenges in LLM Pre-Training}
\label{sec:challenges}

{\bf Term definitions. } To avoid confusion, we define several terms used throughout this paper. {\bf Replica: } one complete copy of the model. {\bf Rank: } the index of a device among a certain group.
{\bf Microbatch: } one forward-backward pass done by a replica. {\bf Gradient synchronization: }
gradient all-reduce across the replicas. {\bf Gradient accumulation: } gradient aggregation of
$G$ microbatches on each replica before gradient synchronization. {\bf Iteration: } the full cycle
of gradient accumulation, synchronization, model update.

\begin{figure}[t]
\centering
\includegraphics[width=\linewidth]{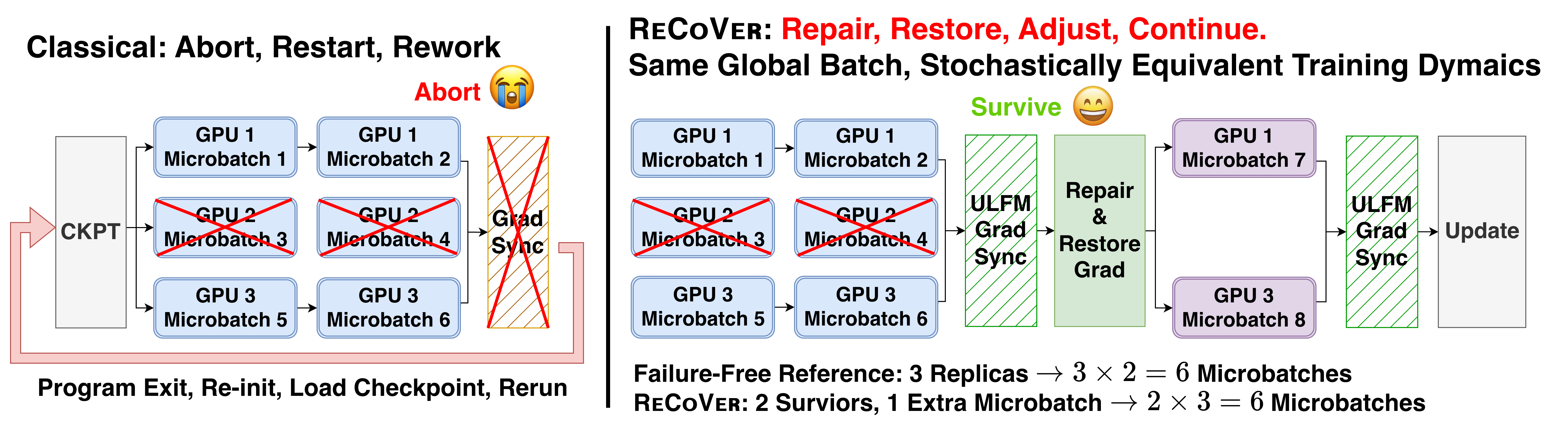}
\vspace{-15pt}
\caption{Comparison of a classical synchronous iteration (left) and \framework (right) under a mid-iteration failure. The classical approach aborts all replicas; \framework localizes and repairs broken gradient sync in place, adjusts survivor workload, and completes with the same microbatch count, preserving stochastically equivalent update.}
\vspace{-10pt}
\label{fig:layout}
\end{figure}

{\bf Iteration anatomy and what happens after failure.} During an iteration, replicas compute microbatches independently and remain unaware of remote failures until gradient synchronization, when failures are
detected. As shown in Figure.~\ref{fig:layout}, traditional checkpoint-restart approaches must then abort all replicas and discard all progress since the last checkpoint, despite failures typically affecting only a small
number of replicas (often just one). We therefore argue that resilience at hyperscale pre-training needs
to protect this synchronization point, such as to be able to contain failures locally to the affected replicas, which is a premise of forward recovery.

{\bf Five challenges.}
\emph{{\color{blue} C1: standard communication backends do not survive rank loss.}} Nvidia Collective Communication Library (NCCL) is the de facto choice for deep learning applications and classical Message Passing Interface (MPI) is the mainstream communication protocol in high-performance-computing (HPC) community, yet they both have no application-facing way to complete an in-flight collective under failure or form a new process group over survivors. Recent works \cite{salpekar2026training,wang2025reliable,ulfm,ulfm-mpich} have made prominent progress, however, none of which have become standard in LLM applications. A clean post-failure communicator (C1) is necessary but not sufficient. \emph{\color{blue} {C2: drop-and-go is not computationally equivalent.}} After satisfying C1, one solution would be to drop the faulty replica and continue with the survivors. However, such approach would use fewer microbatches in the gradient synchronization, shifting the gradient-noise scale and inducing loss spikes that deviate from the baseline
configuration. \emph{\color{blue} {C3: asynchronous rejoin does not solve computational equivalence.}} Even if replicas may rejoin later~\cite{salpekar2026training}, under frequent failures, a degraded mode with
smaller world size would be the norm rather than exception, thus suffering from the same issues as drop-and-go. \emph{\color{blue} {C4: hot spares are rigid and wasteful.}} Using spare replicas addresses
C2--C3 at the cost of resource redundancy, which is expensive for large models that need many GPUs for
each replica. \emph{\color{blue} {C5: parallelism-specific designs lack versatility.}} A versatile protocol must decouple the cross-replica failure-recovery logic from the intra-replica communication structure to allow seamless forward recovery integration.

\section{The \framework Framework}
\label{sec:framework}

\begin{figure}[t]
\centering
\includegraphics[width=\linewidth]{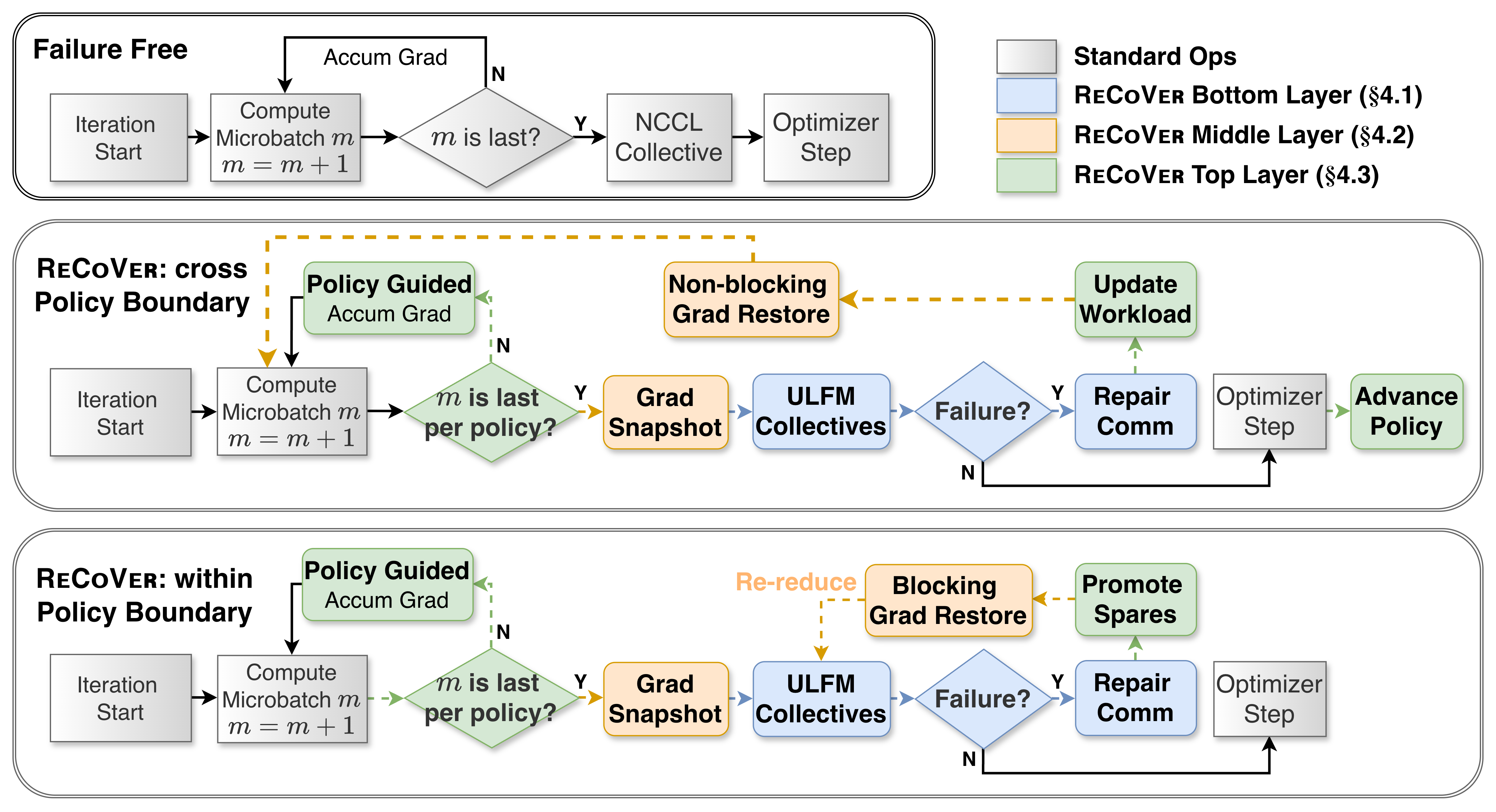}
\vspace{-15pt}
\caption{Flowchart of a \framework iteration and how its three-layer protocol interacts. }
\vspace{-10pt}
\label{fig:workflow}
\end{figure}

\paragraph{Design overview.}
\framework is a three-layer fault-tolerant protocol (overview in \cref{fig:layout}, details in \cref{alg:stepflow-full}). At the bottom layer, a fault-tolerant collective \ulfmallreduce (\cref{sec:ulfm-guarded}) verifies communicator health \emph{before} each all-reduce. Upon failure, it repairs the communicator over surviving ranks and either early-returns or performs a guarded reduction, ensuring no fatal errors and returning
\begin{wrapfigure}{r}{0.5\linewidth}
\vspace{-15pt}
\begin{minipage}{\linewidth}
\begin{algorithm}[H]
\small
\caption{\small A \framework iteration.}
\label{alg:stepflow-full}
\begin{algorithmic}[1]
\REQUIRE policy $P$; replica role $\rho$; policy-assigned per-role workload $P(\rho)$; target total workload $B$; bucket bookkeeping $\mathcal{B}$; communicator epoch $\epsilon_{\mathrm{cur}}$; intra-replica group $PG_{\mathrm{intra}}$; cross-replica group $PG_{\mathrm{cross}}$
\vspace{-8pt}
\STATE $\mathcal{B} \gets \varnothing$;\ zero gradients;\ $m \gets 0$.
\WHILE{microbatch index $m < P(\mathrm{major})$} \label{line:outer-while}
  \STATE $m \gets m + 1$;\ run forward and backward.
  \STATE {\bf if} $m \leq P(\rho)$ {\bf then} locally accumulate gradient;
  \STATE {\bf else} zero this microbatch's gradient.
  \STATE {\bf end if}
  \IF{$m = P(\mathrm{major})$}
      \FORALL{newly produced gradient bucket $b$}
        \STATE snapshot $S(b)\leftarrow b$;\ tag $\epsilon(b) \leftarrow \epsilon_{\mathrm{cur}}$;
        \STATE append $(b, S(b), \epsilon(b))$ to $\mathcal{B}$.
        \IF{$\rho \in \{\text{major-spare, minor-spare}\}$}
            \STATE zero $b$.
        \ENDIF
        \STATE \textbf{post} \callulfmallreduce on $PG_{\mathrm{cross}}$.
      \ENDFOR
      \STATE \textbf{post} NCCL barrier on $PG_{\mathrm{intra}}$.
      \STATE \textbf{post} \callulfmconsensus on $PG_{\mathrm{cross}}$.
      \IF{any ULFM collective returns failure}
        \STATE {\bf call} \callhandlefail.
        \STATE {\bf call} \callgradrestore.
        \STATE $P, \rho \leftarrow $ \callpolicyadjust.
      \ENDIF
  \ENDIF
\ENDWHILE
\STATE divide accumulated gradient by $B$;\ optimizer step.
\IF{crossed a policy boundary}
  \STATE $P, \rho \leftarrow$ \callpolicyadvance.
\ENDIF
\end{algorithmic}
\end{algorithm}
\end{minipage}
\vspace{-30pt}
\end{wrapfigure}
globally consistent status for all callers. The middle layer, \emph{in-step fine-grained recovery} (\cref{sec:instep}), uses this signal to restore gradients to their pre-reduction state, preventing inconsistent reductions and preserve survivor's intra-iteration progress. The top layer, \emph{versatile workload} (\cref{sec:versatile}), a policy that dynamically adjusts replica workloads after failures to maintain a constant per-step microbatch count.

\vspace{-5pt}
\paragraph{Claim.}
Together, the three layers ensure that, as long as one replica survives, each \framework iteration is stochastically equivalent to a failure-free run: every iteration aggregates the same number of microbatch gradients before the optimizer step. Failed replicas’ data partitions are dropped, while survivors process their partitions faster. Given the large pre-training data stream, this is equivalent to a different random shuffle of the same corpus, so per-iteration gradients follow the same distribution as in the failure-free case. \Cref{app:proof} formalizes this.

\subsection{Bottom Layer: ULFM-Guarded Fault-Tolerant Collectives}
\label{sec:ulfm-guarded}

\begin{wrapfigure}{r}{0.32\linewidth}
\centering
\vspace{-20pt}
\includegraphics[width=\linewidth]{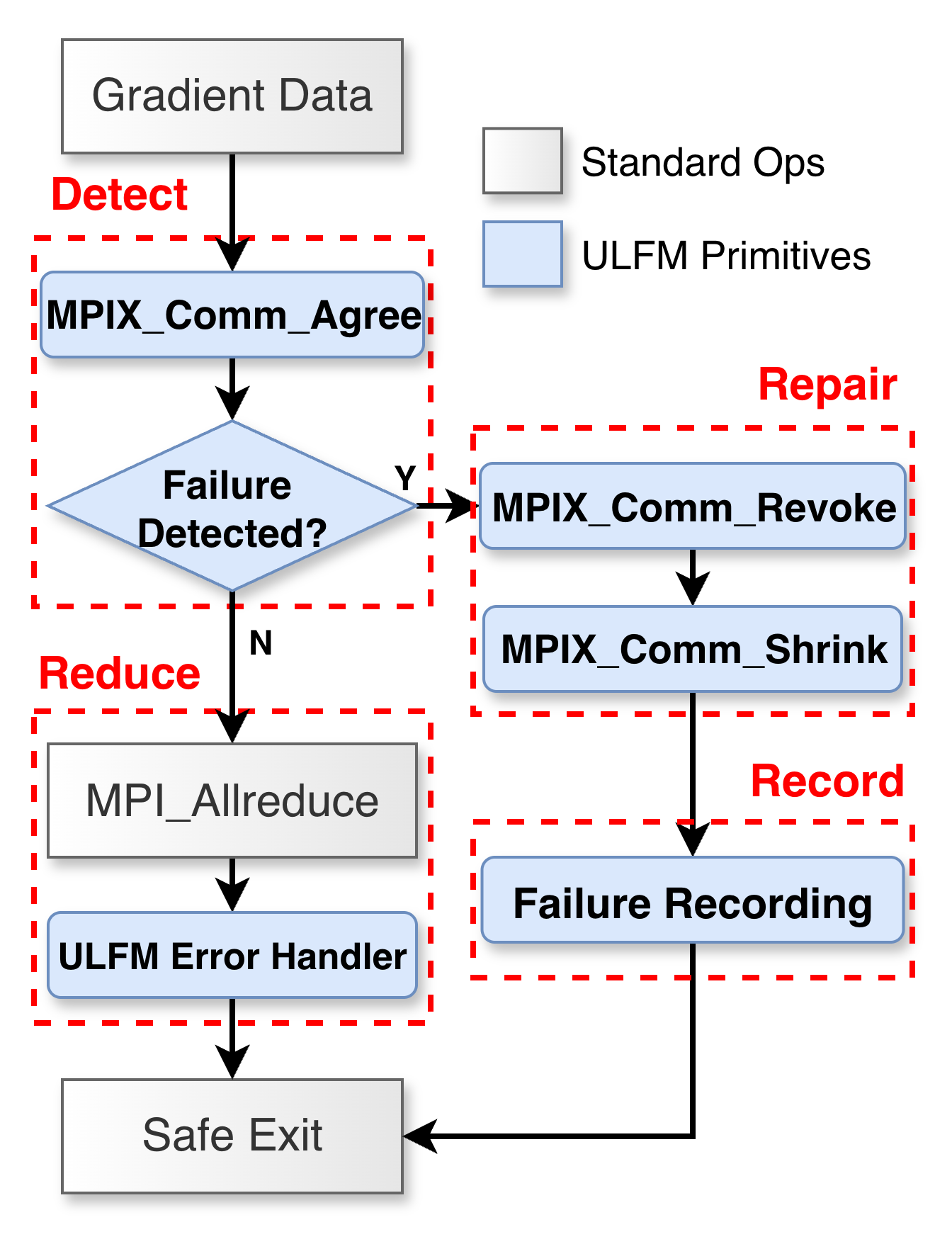}
\vspace{-20pt}
\caption{\small \framework bottom layer.}
\label{fig:bottom-layer}
\vspace{-20pt}
\end{wrapfigure}

\paragraph{Goal and primitives.}
The bottom layer's goal is to provide resilient collective primitives for the rest of the framework to survive rank loss: every caller eventually returns, either completing the reduction on a repaired communicator or surfacing a collectively agreed consistent failure signal. \framework adds two such primitives to PyTorch's backend (\cref{fig:bottom-layer}): \ulfmallreduce, a fault-tolerant all-reduce used wherever the application would issue a standard cross-replica one; and \ulfmconsensus, a barrier-like collective that guarantees the consistency of replicas (e.g., all ranks of a replica either succeed together or discover failures together) and ensures all replicas agree on the same view for upper-layer's recovery logic (\cref{sec:instep,sec:versatile}).

\vspace{-5pt}
\paragraph{Why ULFM is a good foundation.}
User-Level Failure Mitigation (ULFM)~\cite{ulfm} extends MPI with the minimal semantics these primitives need: no MPI call blocks indefinitely after a failure; it either succeeds or returns a typed error. Unlike NCCL or conventional MPI, where any rank loss aborts the job, ULFM exposes failures at the communicator level and leaves recovery path to the application. We build on two of its collective routines: (1) \texttt{MPIX\_Comm\_agree}, which implements fault-tolerant consensus (i.e., returns success only if all communicator members are alive); and (2) \texttt{MPIX\_Comm\_shrink}, which builds a new communicator that includes only surviving members. ULFM is intentionally minimal, focusing solely on collective membership management. Any buffers and data involved in failed collectives are left in an unspecified state. \framework guarantees recovery of these data structures at upper layers, which is only feasible based on the strong foundation of ULFM that keeps the application alive.

\vspace{-5pt}
\paragraph{Building the primitives.}
\ulfmallreduce (\cref{fig:bottom-layer}) composes these routines under a deliberately conservative contract: \emph{avoid issuing an all-reduce on a communicator on which some processes might have already died}. Given a tensor, a reduction operator, and a process group $PG$, it runs four steps:
(1) \emph{Detect.} Call \texttt{MPIX\_Comm\_agree} on $PG$; if successful, jump to step~4;
(2) \emph{Repair.} A failure happened; revoke $PG$'s communicator and call \texttt{MPIX\_Comm\_shrink} to form a communicator over survivors;
(3) \emph{Record.} On the repaired communicator, agree on application-specific failure knowledge and return early;
(4) \emph{Reduce.} Issue the MPI all-reduce on the validated communicator.
Similarly, \ulfmconsensus follows steps 1--3 for a barrier-like failure detection and recovery primitive.


{\bf Why these primitives address \emph{C1}.} Failures are safely isolated to faulty replicas (right half of~\cref{fig:layout}), and the upper layers choose how to proceed instead of being forced to abort. This also lays the groundwork for \emph{C5}: communication resilience is exposed as a drop-in collective interface, independent from any specific intra-replica communication structure, so any layer above can adopt it without changing its parallelism. The remaining challenges \emph{C2--C4} and the full decoupling needed for \emph{C5} are addressed by the upper layers under LLM pre-training settings.

\subsection{Middle Layer: In-Step Fine-Grained Failure Recovery}
\label{sec:instep}

\begin{wrapfigure}{r}{0.35\linewidth}
\centering
\vspace{-25pt}
\includegraphics[width=\linewidth]{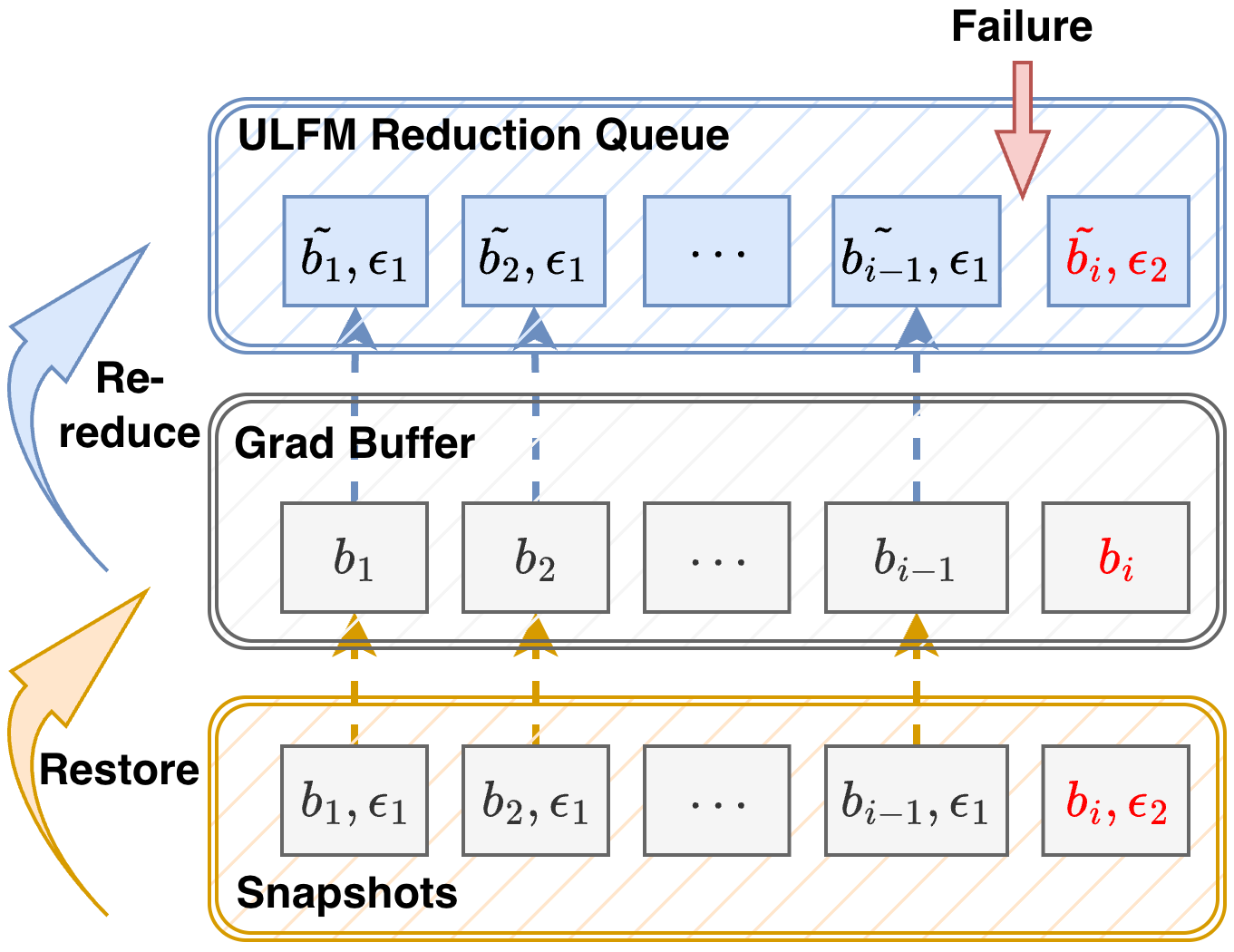}
\vspace{-15pt}
\caption{\small \framework middle layer}
\label{fig:middle-layer}
\vspace{-20pt}
\end{wrapfigure}

\paragraph{Why fine-grained recovery matters at scale.}
A single LLM pre-training iteration spans many microbatches, yielding global batches of hundreds to thousands of millions of tokens that scale with total training tokens \cite{zhang2024does}. This regime is not only viable but desirable, as it amortizes latency-bound cross-replica communication and improves GPU utilization. Consequently, one iteration may last minutes, while failures occur every tens of minutes on a 100K-GPU cluster, making it essential to preserve progress within each iteration.

\vspace{-5pt}
\paragraph{Failure anatomy inside an iteration.}
 Failures can occur in three cases: {\bf (a) Before gradient synchronization.} Buckets contain only local gradients; the first \ulfmallreduce uses a shrunk communicator, so no reductions span different memberships. {\bf (b) After synchronization, before the optimizer step.} All reductions complete in the original world, so gradients remain valid and the failure can be handled in the next iteration safely. {\bf (c) During synchronization.} Some gradients were reduced in the original world and include contributions from replicas absent in the shrunk world.  Above-iteration schemes like FTAR \cite{salpekar2026training} must discard progress, whereas \framework avoids this limitation via ULFM’s communicator-level fault tolerance and a per gradient-bucket bookkeeping.

{\bf Per gradient-bucket bookkeeping.}
Before each gradient all-reduce, \framework snapshots the bucket’s pre-reduce state along with a monotonic \emph{world epoch} that increments on each communicator shrink. Upon failure, only snapshots from smaller \emph{world epoch} are restored. This fine-grained tracking is enabled by ULFM’s communicator-level failure detection and recovery, preserving the process group context—unlike \cite{salpekar2026training}, which requires tearing down and rebuilding the group.

{\bf Blocking and non-blocking restoration.}
\framework selects between two restoration strategies based on whether the iteration proceeds to the optimizer step or defers to a policy boundary (\cref{sec:versatile}). (1) \emph{Blocking restoration} synchronously restores corrupted buckets and re-reduces them under the shrunk world before the optimizer step. (2) \emph{Non-blocking restoration} schedules the snapshot rewind on a separate CUDA stream and overlaps it with the forward pass of the first extra microbatch at the policy boundary step. This step issues a fresh cascade of all-reduces, eliminating the need to manually re-reduce corrupted buckets.  The next subsection's policy layer decides between the two.

\subsection{Top Layer: Versatile-Workload for Training Trajectory Preservation}
\label{sec:versatile}

{\bf Workload redistribution and justification.}
 The versatile workload relies on two properties of gradient accumulation. \emph{(i) Sum-level fungibility:} an iteration’s gradient is the sum of $B$ microbatch gradients, independent of which replica computes each term, any partition of $B$ microbatches across survivors is equivalent. \emph{(ii) Stream-level exchangeability:} when a replica fails, replacement microbatches are drawn from the survivors' own partitions of  an effectively infinite, exchangeable data stream, yielding a $B$-batch that is distributionally equivalent to the original (\cref{app:proof}). Property (i) guides redistribution; property (ii) ensures it preserves the training trajectory. \framework leverages both by assigning each survivor replica one of four roles per iteration.


{\bf Four replica roles and invariant.} {\bf Major:} contributes $G_{\mathrm{cur}}$ microbatches.
 {\bf Minor:} contributes $R_{\mathrm{cur}} < G_{\mathrm{cur}}$ microbatches to absorb any remainder when $W_{\mathrm{cur}} \times G_{\mathrm{cur}} > B$, with $W_{\mathrm{cur}}$ being the current replica count. {\bf Major-/minor-spare:} executes the same workload as its counterpart but zeros its gradient buffer at all-reduce; it does not impact the global batch until promoted after a failure. Let $C_r(t)$ be the number of microbatches replica $r$ contributes at iteration $t$ (zero for spares). \framework targets a single invariant across all iterations:
\begin{equation}
\sum_{r \in \text{survivors}(t)} C_r(t) \;=\; W_{\mathrm{init}} \cdot G_{\mathrm{init}} \;=\; B.
\label{eq:invariant}
\end{equation}
We call the family of replica-role assignments that maintains
\cref{eq:invariant} the \emph{versatile workload}.

\begin{figure}[t]
\centering
\includegraphics[width=\linewidth]{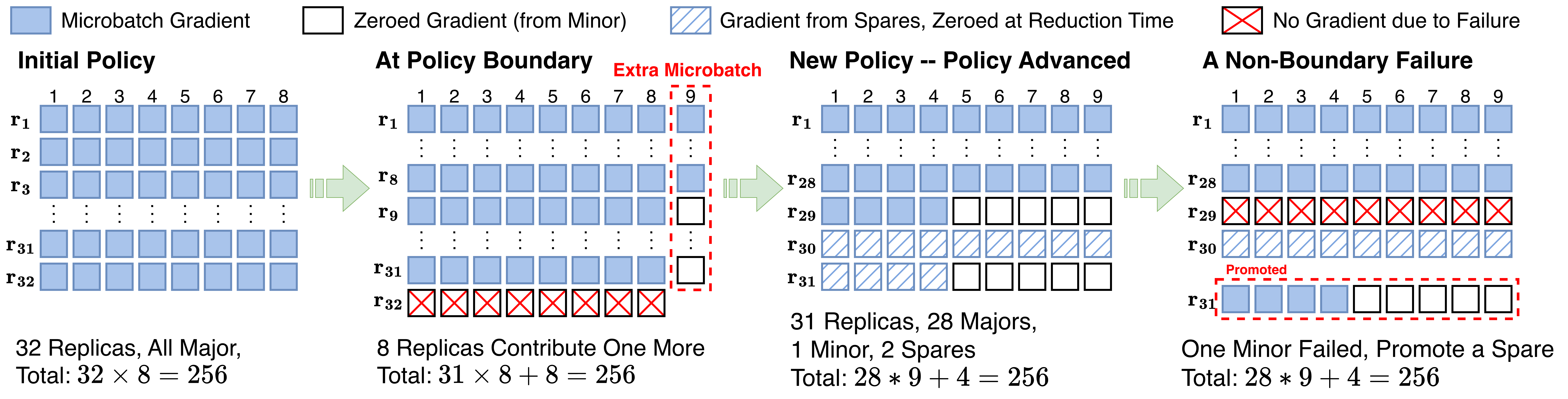}
\vspace{-15pt}
\caption{Versatile workload across two failures. (i)
Pre-failure: all replicas are major. (ii) First failure crosses policy boundary: no spares available, compute extra microbatches. (iii) Policy advanced: $28$ majors, $1$ minor, and $2$ spares. (iv) Second failure within policy boundary: spares available, promote.}
\vspace{-10pt}
\label{fig:workload}
\end{figure}

{\bf Versatile-workload policy.} At each iteration, replica roles adapt to failures based on whether a {\bf Policy Boundary} is reached, i.e., whether available spares can replace failed replicas. If so, spares are promoted to majors/minors and the current accumulation $G_{\mathrm{cur}}$ is unchanged. Otherwise, Eq.~\eqref{eq:invariant} no longer holds unless a {\bf Policy Boundary Step} being performed. Specifically, for a survivor count $W_{\mathrm{cur}}$, the collectively agreed total contribution from survivors $C_{\mathrm{cur}}=\sum_{r} C_r(t)$ (obtained from phase 3 in \ulfmallreduce), and the extra grad-accum steps $G_{\mathrm{ext}}=1$, the policy increments $G_{\mathrm{ext}}$ until $C_{\mathrm{cur}} + W_{\mathrm{cur}} \cdot G_{\mathrm{ext}} \geq B$. In case of an inequality, assign $C_{\mathrm{cur}} + W_{\mathrm{cur}} \cdot G_{\mathrm{ext}} - B$ replicas as {\bf boundary minors}, a temporary role that each of them contributes $G_{\mathrm{ext}}-1$ extra microbatches. After committing the curren iteration, the policy increments $G_{\mathrm{cur}}$ until $W_{\mathrm{cur}} G_{\mathrm{cur}} \geq B$, then sets $N=\lfloor B / G_{\mathrm{cur}} \rfloor$. If $N \cdot G_{\mathrm{cur}} = B$, assign $N$ majors and the rest as major-spares; otherwise assign $N$ majors, one minor with $R_{\mathrm{cur}} = B - N\cdot G_{\mathrm{cur}}$, and distribute remaining replicas as spares.


{\bf Illustrative example.}
\Cref{fig:workload} instantiates the policy for $W_{\mathrm{init}} =
32$ and $G_{\mathrm{init}} = 8$, so $B = 256$. The four panels trace
the policy through one boundary step and a subsequent non-boundary
failure.

\emph{Pre-failure.} All $32$ replicas are majors at $G_{\mathrm{init}} = 8$, and $32 \times 8 = 256 = B$.

\emph{A replica fails during iteration $t$.} Now $W_{\mathrm{cur}} = 31$ and $C_{\mathrm{cur}} = 31 \cdot 8 = 248$, no spares so it's a policy boundary step: eight survivors contribute one extra microbatch giving $248 + 8 = 256 = B$.

\emph{Policy advancement.} Incrementing $G_{\mathrm{cur}}$ by 1 yields $31\times 9=279 > 256 = B$, so $G_{\mathrm{cur}}=9$. We have $N = \lfloor 256/9 \rfloor = 28$ and $28 \cdot 9 = 252 \neq 256$, so the assignment is $28$ majors at 9 microbatches and one minor at $R_{\mathrm{cur}} = 256 - 252 = 4$ microbatches, with the one of remaining $31 - 29 = 2$ survivors as major spare, one as minor spare.

\emph{Next failure, spare available.} Each spare runs same microbatches as its non-spare counterparts but has its gradients zeroed at all-reduce time. When the minor $r_{29}$ fails mid-window, the minor spare $r_{31}$ is promoted into the vacated role simply by restoring its gradients (\cref{sec:instep}).

\subsection{Integration with LLM Pre-Training Frameworks}
\label{sec:integration}

\begin{wrapfigure}{r}{0.35\linewidth}
\centering
\vspace{-20pt}
\includegraphics[width=\linewidth]{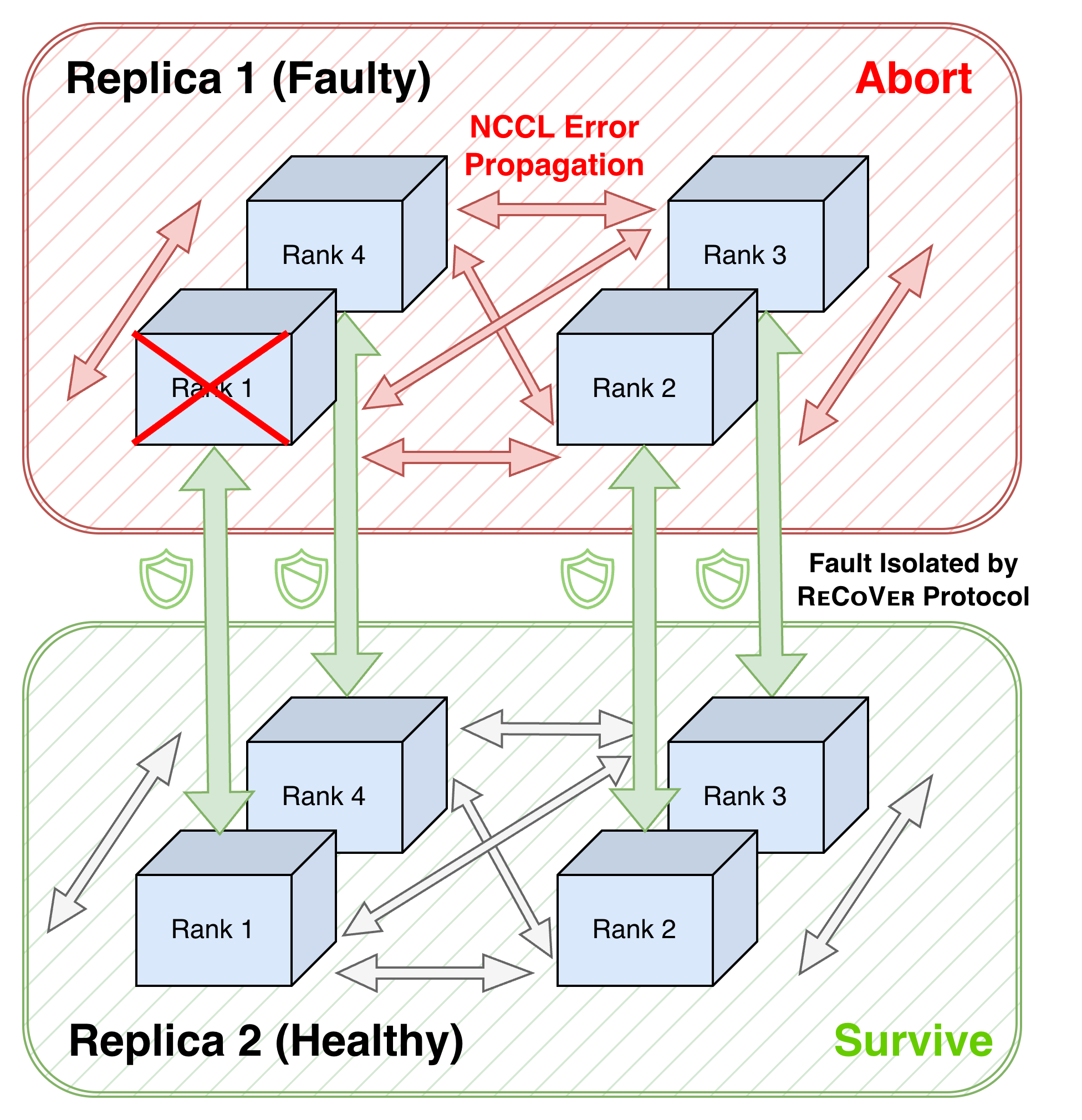}
\vspace{-10pt}
\caption{\small \framework integration with model parallelism}
\label{fig:recover-mp}
\vspace{-20pt}
\end{wrapfigure}
A realistic LLM pre-training system distributes one replica across many devices, each holding a distinct shard of parameters and gradients. \framework naturally lifts onto this setting: every intra-replica rank fires the cross-replica all-reduce and runs the protocol in lockstep. Therefore,
\framework is agnostic to replica internals and \emph{versatile} across parallelism schemes. However, the inner structure introduces nuances, which we address for {\bf 3D parallelism (Data, Tensor, Pipeline)} and {\bf Hybrid Sharded Data Parallel (HSDP)} below.

\vspace{-7pt}
\paragraph{Where the two parallelism substrates agree.}
In both 3D parallelism and HSDP, losing any rank equals to losing an essential shard of the model, thus invalidating the entire replica, making the replica the atomic survival unit under \framework. We keep NCCL for intra-replica communication, while deploying the collectives of \cref{sec:ulfm-guarded} for cross-replica communication. We post an additional replica-consistency gate via a NCCL barrier so that any rank failure will eventually trigger NCCL watchdog timeouts on all intra-replica peers and aborting the replica as a unit. Surviving replicas remain unaffected until the failure detected by their \ulfmallreduce or \ulfmconsensus, which then repairs and shrinks the communicator to surviving replicas. The remainder of the iteration is then driven by the recovery and policy mechanisms of \cref{sec:versatile,sec:instep}. \Cref{fig:recover-mp} illustrates this asymmetry between intra-replica abortion and cross-replica survival.

\vspace{-7pt}
\paragraph{Where the two substrates differ.}
The integrations differ in three minor aspects.
\emph{(i)~ULFM-guarded cross-replica group.} \framework-3D guards the data-parallel all-reduce, whereas \framework-HSDP protects the cross-\emph{replicate} group all-reduce.
\emph{(ii)~Snapshots.} \framework-3D snapshots bucketed gradients of its local shard, while \framework-HSDP snapshots sharded gradients of flattened FSDP parameters.
\emph{(iii)~Replica-consistency gate.} \framework-3D places the NCCL barrier on the union of TP and PP groups, whereas \framework-HSDP barriers on the FSDP shard group.

\section{Evaluation}
\label{sec:eval}

We evaluate \framework on the 3D-parallel integration (\framework-3D) of \cref{sec:integration}, addressing three questions: (i)~does it preserve the failure-free optimization trajectory under frequent failures? (ii)~what resource overhead does this preservation incur over an ideal never-failing scenario? and (iii)~how does it compare to a standard checkpoint-restart~\cite{nanotron,paszke2019pytorch} approach under single and consecutive failures? We argue that comparing with the failure-free reference is not just enough, but the most challenging setting: a resilient system can do at most as good as its never-failing counterpart. And we only compare with standard checkpoint because they are complementary to \framework: an advanced checkpointing method can be combined with our \emph{forward recovery} system to further boost resilience. These settings are also used in the most recent production effort~\cite{salpekar2026training}, which \framework is conceptually complementary with and can be engineered to be combined together. The HSDP integration (\framework-HSDP) uses the same setting and results are presented in \cref{app:hsdp-eval}.

\begin{figure}[t]
\centering
\begin{subfigure}[t]{0.47\linewidth}
\centering
\includegraphics[width=\linewidth]{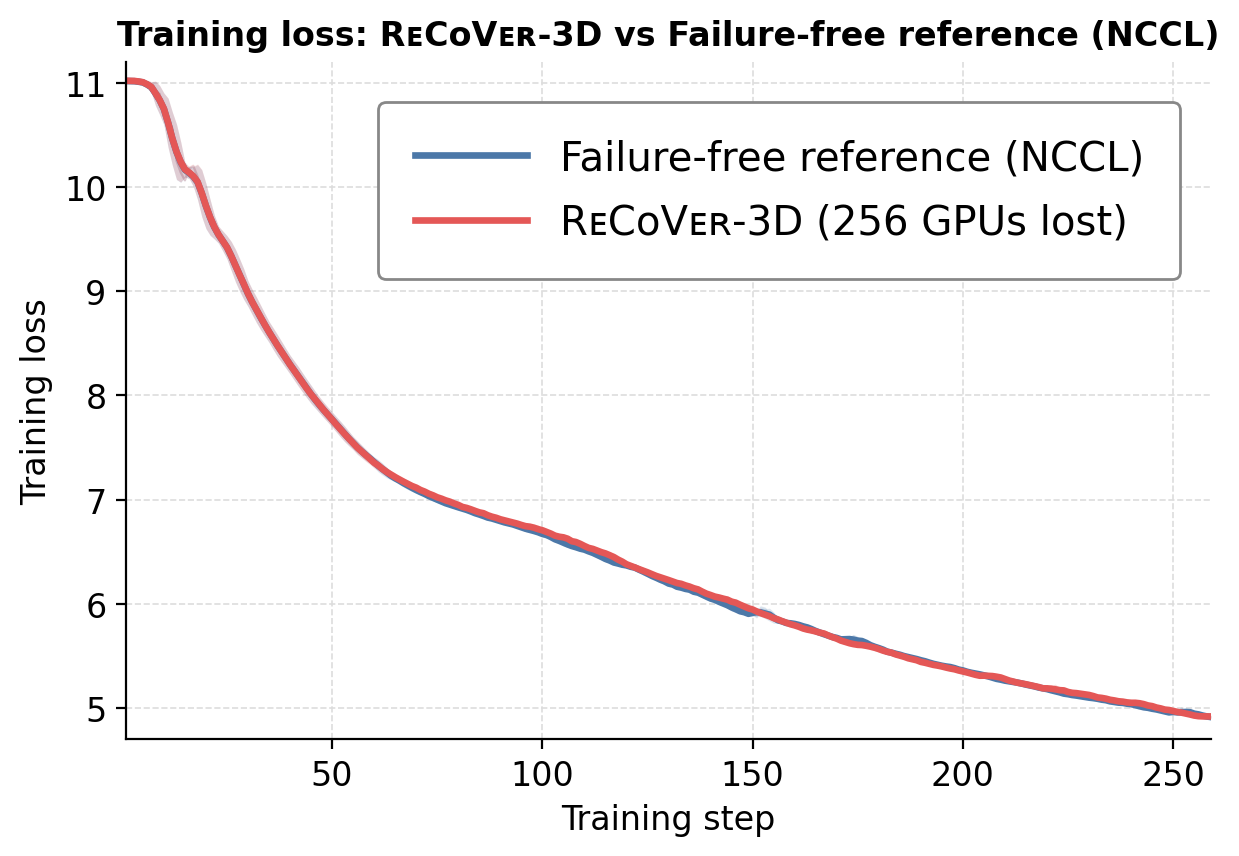}
\caption{}
\label{fig:eval-trajectory-loss}
\end{subfigure}\hfill
\begin{subfigure}[t]{0.53\linewidth}
\centering
\includegraphics[width=\linewidth]{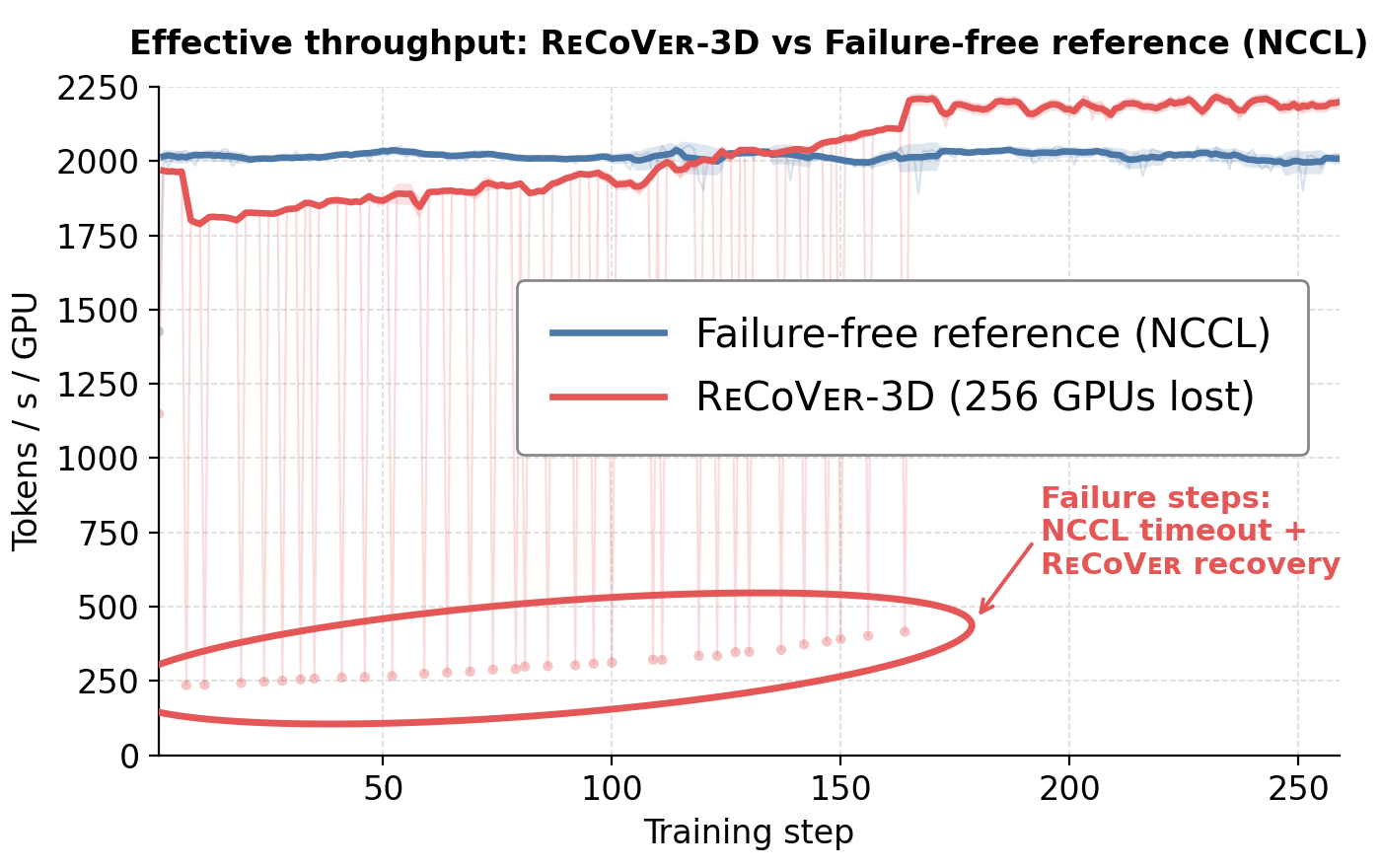}
\caption{}
\label{fig:eval-trajectory-throughput}
\end{subfigure}
\caption{Trajectory preservation under $256$ GPU losses on a 512-GPU 3D-parallelism run. \subref{fig:eval-trajectory-loss}~The
\framework-3D training loss curve matches the failure-free
NCCL reference, with no spikes or measurable deviation. \subref{fig:eval-trajectory-throughput}~\framework-3D improves per-GPU utilization along the failures due to versatile workload thus surpassing the failure-free reference.}
\vspace{-15pt}
\end{figure}

{\bf Setup.} We pre-train a 7B-parameter LLaMA-style model on C4~\cite{raffel2020c4} dataset under a 3D
parallelism stack (TP$\times$PP$\times$DP), running on 512 A100 40GB GPUs, with the initial replica count
$W_{\mathrm{init}}=64$ and grad-accum factor $G_{\mathrm{init}}=128$ given a global $B = W_{\mathrm{init}} \cdot G_{\mathrm{init}} = 8192$ microbatches per optimizer step. All failures are injected on a randomly chosen GPU via a randomly generated deterministic schedule by the simulator of \cref{app:impl}: $256$ GPU losses are spread across the run, spaced at every 5 iterations, each at a randomly chosen step within its interval, to stress-test the system under frequent failure regimes. We deliberately inject all failures \emph{\color{blue}during gradient synchronization}, which is the most difficult setting due to the existence of partially reduced gradients. We use grad-accum $=8$ when comparing to \emph{checkpoint-restart baseline} so that every component in the break-down of a failure recovery window would remain comparable. For fair comparisons, we sweep the baseline's checkpoint interval $N$ from $2$ to $64$ iterations, and inject failures right at the middle of each interval. This reflects the practical consideration that checkpoint frequency should adapt to failure frequency. Our headline metric is \emph{effective throughput} (tokens per second per alive GPU): unlike raw throughput, it factors out the shrinking world and directly measures resource utilization. Full evaluation details are in \cref{app:eval-details}.

{\bf Trajectory preservation: indistinguishable curves.}
\Cref{fig:eval-trajectory-loss} plots the training loss of
\framework-3D against the failure-free NCCL reference over the
first $260$ optimizer steps. Despite $256$ GPU losses spread
across the run, the two curves are visually indistinguishable, with
no spikes, oscillations, or post-failure correction transients.
This is the empirical evidence of the trajectory-preservation
guarantee of \cref{app:proof}: every iteration commits the same number of microbatches to the optimizer, so the per-iteration gradient distribution is the same as the reference's, and the optimization trajectory inherits that equivalence. By contrast, prior keep-alive frameworks that allow the global batch to contract between failures and reconfigurations have been observed to produce loss fluctuations contingent on the failure schedule~\cite{salpekar2026training}; \framework intrinsically removes that contingency.

{\bf Quantitative equivalence and data coverage.} Over the first $250$ steps, \framework-3D consumes the same $8.39$B tokens as the reference, of which $2.79$B ($33\%$) are drawn from survivors' streams in place of the shards held by the $32$ lost replicas. No data is lost, only deferred: the displaced indices are determined by when and which replica failed, and can be scheduled into a subsequent run. As each replica's shard is an unbiased sample of the data mixture (Nanoset~\cite{nanotron} randomly permutes the weighted global sample index before striding it across DP ranks), replacing a lost shard is equivalent to reshuffling the data stream under a different seed. Despite $33\%$ of the tokens differing, the two runs remain statistically identical: the validation loss differs by at most $0.024$ ($0.36\%$) and ends $0.008$ lower (\cref{tab:val-loss}). The same holds for \framework-HSDP (mean training-loss deviation $0.009$, \cref{app:hsdp-eval}) and under i.i.d.\ per-GPU failures (\cref{app:iid-failures}).

\begin{table}[H]
\centering
\small
\begin{tabular}{lccccc}
\toprule
Step & 50 & 100 & 150 & 200 & 250 \\
\midrule
Failure-free reference (NCCL) & 7.723 & 6.714 & 5.922 & 5.342 & 4.961 \\
\framework-3D ($256$ GPUs lost) & 7.721 & 6.689 & 5.923 & 5.342 & 4.953 \\
\bottomrule
\end{tabular}
\vspace{5pt}
\caption{Validation loss on held-out data for the 512-GPU 3D-parallelism run of \cref{fig:eval-trajectory-loss}.}
\label{tab:val-loss}
\end{table}

{\bf Effective throughput: drop, climb, exceed.}
\Cref{fig:eval-trajectory-throughput} shows effective throughput along the same run. Pre-failure, \framework-3D matches the NCCL
reference closely, where the gap comes from two parts: (1) we implemented \framework without overlapping gradient synchronization with backward computation, while the baseline does; (2) the backend overhead of OpenMPI over NCCL. At the first failure, throughput drops
sharply, which we speculate that it is caused by the cross-replica world reshape producing a topology that is less efficient for MPI backend. As more failures occur, gradient accumulation increases and amortizes the backend overhead by adding more compute per iteration, effectively raising per-GPU utilization. Concretely, with $G$ the current grad-accum factor, $s$ the tokens per microbatch, $t_c$ the compute time per microbatch, and $T_r$ the per-iteration cross-replica all-reduce cost, which stays constant as the gradient size is fixed by the model, the per-GPU effective throughput is 
$$G s/(G t_c + T_r) = s/(t_c + T_r/G),$$ 
monotonically increasing in $G$ toward the pure-compute rate $s/t_c$. The NCCL reference obeys the same form at its fixed $G_{\mathrm{init}}$, so \framework-3D overtakes it once $$T_r^{\mathrm{MPI}}/G_{\mathrm{cur}} < T_r^{\mathrm{NCCL}}/G_{\mathrm{init}}.$$
In the high-failure regime the amortization dominates, and \framework-3D's effective
throughput climbs back and eventually exceeds the failure-free reference. This is
not an artifact of the metric: each surviving GPU does
\emph{strictly more useful compute per unit time}, even as the
optimization trajectory remains stochastically equivalent to the
failure-free run.

\begin{figure}[t]
\centering
\begin{minipage}[c]{0.52\linewidth}
\centering
\begin{subfigure}[t]{\linewidth}
\centering
\vspace{10pt}
\includegraphics[width=\linewidth]{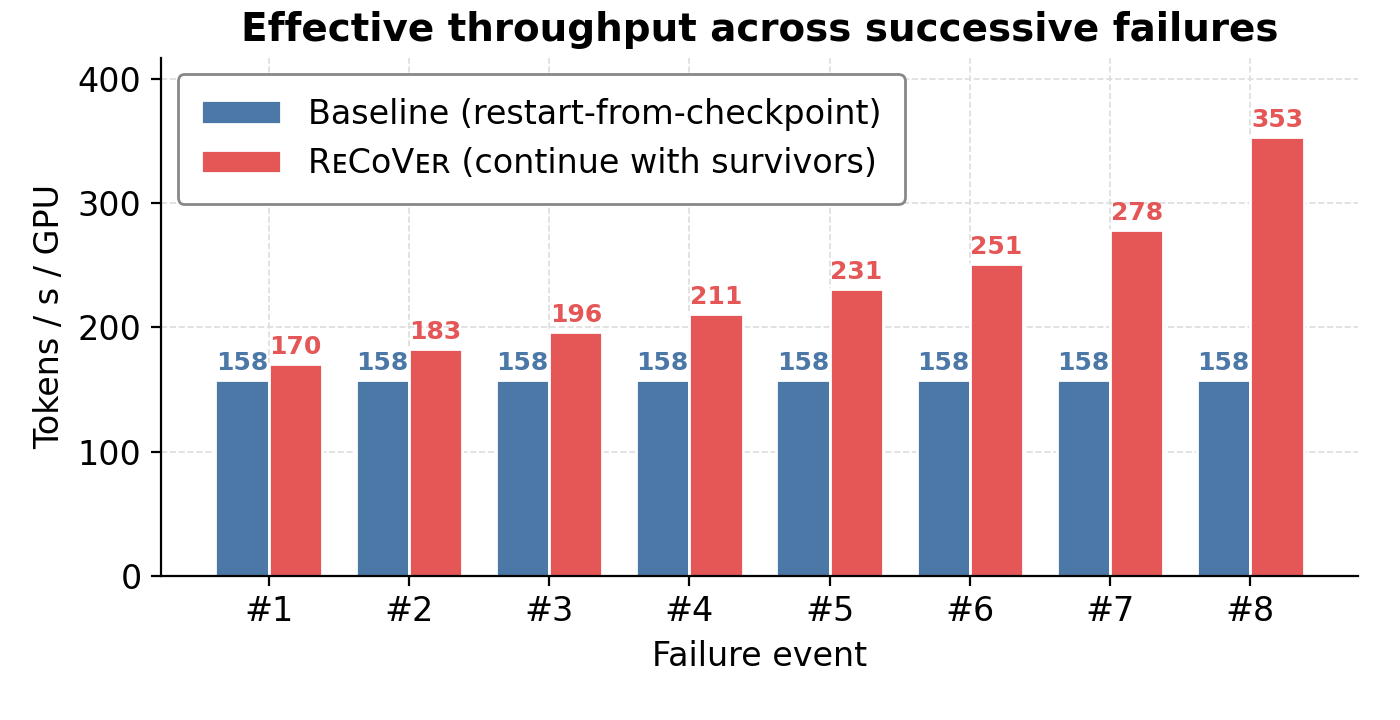}
\vspace{-20pt}
\caption{}
\label{fig:eval-cost-throughput}
\end{subfigure}\\[4pt]
\begin{subfigure}[t]{\linewidth}
\centering
\vspace{2.5pt}
\includegraphics[width=\linewidth]{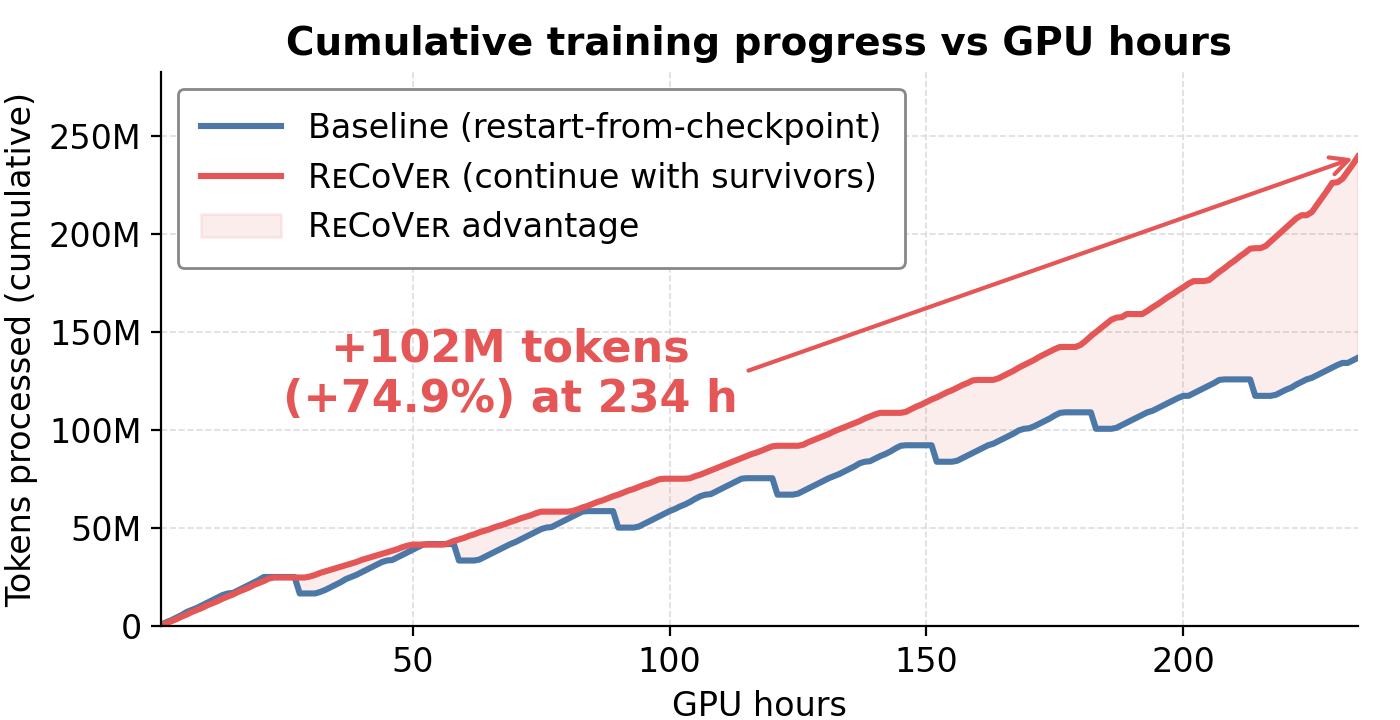}
\vspace{-12pt}
\caption{}
\label{fig:eval-cost-progress}
\end{subfigure}
\end{minipage}\hfill
\begin{subfigure}[c]{0.48\linewidth}
\centering
\includegraphics[width=\linewidth]{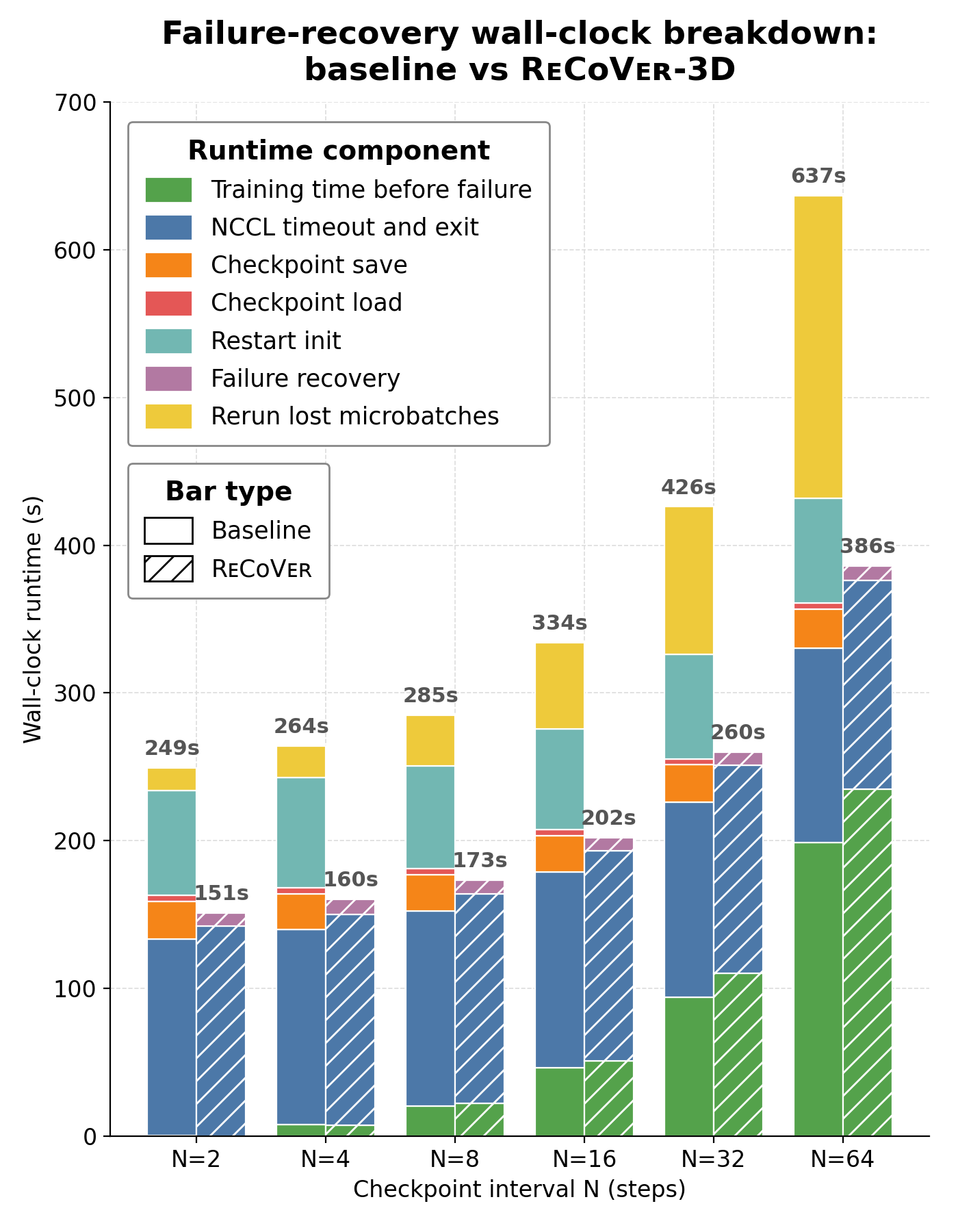}
\vspace{-20pt}
\caption{}
\label{fig:eval-cost-breakdown}
\end{subfigure}
\caption{Cost comparison between \framework-3D and
restart-from-checkpoint. \subref{fig:eval-cost-throughput}~Effective
throughput across successive failures; \framework-3D keeps increasing and enlarges the gap between baseline as the growing per-GPU workload amortizes the cross-replica all-reduce cost. \subref{fig:eval-cost-progress}~Cumulative training progress in tokens vs GPU-hours; \framework-3D processes $74.9\%$ more tokens at $234$ GPU-hours.
\subref{fig:eval-cost-breakdown}~Single-failure raw wall-clock breakdown swept over checkpoint interval $N$.}
\vspace{-10pt}
\label{fig:eval-cost}
\end{figure}

{\bf Comparison with checkpoint-restart, across single and multiple failures.} We now demonstrate how \framework saves GPU hours over traditional checkpoint-restart approach on three fronts. \Cref{fig:eval-cost-throughput} reports
effective throughput across successive failure intervals: the
baseline is flat because every failure pays
the same restart-and-rerun cost in expectation, while
\framework-3D's effective throughput \emph{rises} monotonically as each new failure increases the compute workload for survivors and amplifies the amortization effect, which is then translated to strictly better resource utilization on every alive GPU. \Cref{fig:eval-cost-progress} signifies this into cumulative training progress: at $234$ GPU-hours, \framework-3D has processed $+102$\,M more tokens than the baseline, a $74.9\%$ advantage that grows across the run and will keep growing until all replicas have failed. \Cref{fig:eval-cost-breakdown} decomposes a single recovery into its \emph{raw} wall-clock components and sweeps the checkpoint interval $N$ from 2 to 64 steps. The baseline's recovery cost grows with $N$ because longer checkpoint
intervals translate directly into more lost work to re-execute; \framework's cost is roughly flat across $N$ because it never discards work. \framework even wins at the baseline's most favorable $N$ ($N=2$, extremely frequent checkpoints). Moreover, the restart cost (resource allocation, init, loading, first-step cold-start), though not significant at our test scale, is reported to be $\sim10$ minutes on 100k production system~\cite{salpekar2026training} despite being optimized by industrial engineers. With the rapid scaling trend, the restart time alone will dominate baseline's recovery overhead and the gaps in \cref{fig:eval-cost} will be even wider.

\section{Discussion and Limitations}
\label{sec:discussion}

{\bf Dependence on ULFM and portability to NCCL.} What \framework fundamentally relies on is not MPI itself, but an agreement among the survivors on which ranks have failed. Without such an agreement, survivors may post mismatched collectives and hang the job. NCCL can already rebuild a communicator over the survivors through \texttt{ncclCommShrink}, but it only repairs the communicator and leaves the decision on which ranks to exclude to the application. \framework makes this decision with \texttt{MPIX\_Comm\_agree}, whereas on NCCL it would require an out-of-band agreement mechanism. Since the middle and top layers of \framework never call MPI, they can be ported unchanged to any backend that provides both repair and agreement. Such a port would remove the backend overhead of OpenMPI over NCCL while preserving computational equivalence.

{\bf Replica as the unit of recovery, without rejoin.} When a rank fails, \framework removes its entire replica rather than the failed rank alone. The remaining ranks of that replica no longer have enough GPUs for its original TP and PP configuration, and running them in a different configuration would affect all cross-replica collectives. Supporting such heterogeneous replicas is beyond the scope of this work, although our design does not preclude it. Moreover, removed replicas do not rejoin, so the number of replicas only decreases as failures accumulate. We deliberately let the failed replica exit as a whole in order to keep training progressing, and its resources are thus released immediately. In principle, rejoining can be implemented with \texttt{MPI\_Comm\_spawn}, similar to the rejoin path of FTAR~\cite{salpekar2026training}. However, spawned processes are not yet fully supported by current ULFM implementations. This is a limitation of the backend rather than of \framework, and our design is compatible with rejoining once the backend allows it.

{\bf Comparison with recent systems.} ElasWave~\cite{kang2025elaswave} makes communicator rebuilding cheap, but it recovers only at a step boundary and does not fully disclose how the survivors reach that boundary from an all-reduce in which a rank has failed. \framework detects the failure inside that all-reduce and preserves the survivors' intra-iteration progress, whereas ElasWave keeps only a per-step optimizer snapshot and thus re-runs the interrupted iteration. Morphlux~\cite{kumar2026reconfigurable} optically rewires a reserved spare accelerator into the position of the failed one and then restarts from the latest checkpoint, whereas \framework reserves no spare hardware and completes the failing iteration with the survivors.

\section{Conclusion}
\label{sec:conclusion}

\framework reframes resilient LLM pre-training as a \emph{forward recovery} problem governed by a single invariant: each iteration commits gradients from the same number of microbatches as its failure-free reference, regardless of when or where failures occur. Its three-layer protocol enforces this invariant by locally containing failures at the communication layer, recovering partially reduced gradients within the failing iteration, and dynamically redistributing microbatch quotas across survivors. The design \emph{extends} resilience across the full pre-training stack rather than only the communication primitive, remains \emph{versatile} across both 3D and HSDP parallelism, and preserves \emph{computational equivalence} to the failure-free trajectory without rollback, replay, or pre-allocated idle replicas. Compared to checkpoint-restart baselines, \framework achieves up to $2.23\times$ higher effective throughput after successive failures and processes $74.9\%$ more tokens within 234 GPU-hours, with the advantage growing over longer training runs.
Given modern contemporary HPC systems of +100k-GPUs that are used for pre-training, \framework will not suffer from the increasing restart overhead that becomes a major bottleneck for traditional checkpoint-restart approaches, and remains a resource-efficient resilience solution. In future work, we will refine \framework to recycle the surviving ranks of failed replicas (currently discarded) in order to further improve system throughput. Also, we will explore how to
allow fresh replicas to rejoin dynamically in order to better control the training progress.


\begin{ack}

This material is based upon work supported by
the U.S.\ Department of Energy, Office of Science, Office of Advanced
Scientific Computing Research, Artificial Intelligence for Science program, under contracts DE-SC0025390 and DE-AC02-06CH11357.

This research used resources of the National Energy Research Scientific Computing Center, a DOE Office of Science User Facility
supported by the Office of Science of the U.S. Department of Energy under Contract No. DE-AC02-05CH11231 using NERSC award ASCR-ERCAP0030039, as well as NERSC award ALCC-ERCAP0031379.
\end{ack}

\bibliography{neurips_2026}

@article{salpekar2026training,
  title={Training LLMs with Fault Tolerant HSDP on 100,000 GPUs},
  author={Salpekar, Omkar and Varma, Rohan and Yu, Kenny and Ivanov, Vladimir and Wang, Yang and Sharif, Ahmed and Si, Min and Xu, Shawn and Tian, Feng and Zheng, Shengbao and others},
  journal={arXiv preprint arXiv:2602.00277},
  year={2026}
}

@inproceedings{gandhi2024recycle,
  title={Recycle: Resilient training of large dnns using pipeline adaptation},
  author={Gandhi, Swapnil and Zhao, Mark and Skiadopoulos, Athinagoras and Kozyrakis, Christos},
  booktitle={Proceedings of the ACM SIGOPS 30th Symposium on Operating Systems Principles},
  pages={211--228},
  year={2024}
}

@inproceedings{jang2023oobleck,
  title={Oobleck: Resilient distributed training of large models using pipeline templates},
  author={Jang, Insu and Yang, Zhenning and Zhang, Zhen and Jin, Xin and Chowdhury, Mosharaf},
  booktitle={Proceedings of the 29th Symposium on Operating Systems Principles},
  pages={382--395},
  year={2023}
}

@inproceedings{thorpe2023bamboo,
  title={Bamboo: Making preemptible instances resilient for affordable training of large {DNNs}},
  author={Thorpe, John and Zhao, Pengzhan and Eyolfson, Jonathan and Qiao, Yifan and Jia, Zhihao and Zhang, Minjia and Netravali, Ravi and Xu, Guoqing Harry},
  booktitle={20th USENIX Symposium on Networked Systems Design and Implementation (NSDI 23)},
  pages={497--513},
  year={2023}
}

@inproceedings{li2023elastic,
  title={Elastic deep learning through resilient collective operations},
  author={Li, Jiali and Bosilca, George and Bouteiller, Aurelien and Nicolae, Bogdan},
  booktitle={Proceedings of the SC'23 Workshops of the International Conference on High Performance Computing, Network, Storage, and Analysis},
  pages={44--50},
  year={2023}
}

@article{wang2025boost,
  title={BOOST: BOttleneck-Optimized Scalable Training Framework for Low-Rank Large Language Models},
  author={Wang, Zhengyang and Liu, Ziyue and Zhang, Ruijie and Maurya, Avinash and Hovland, Paul and Nicolae, Bogdan and Cappello, Franck and Zhang, Zheng},
  journal={arXiv preprint arXiv:2512.12131},
  year={2025}
}

@article{lee2026spare,
  title={SPARe: Stacked Parallelism with Adaptive Reordering for Fault-Tolerant LLM Pretraining Systems with 100k+ GPUs},
  author={Lee, Jin and Chen, Zhonghao and He, Xuhang and Underwood, Robert and Nicolae, Bogdan and Cappello, Franck and Lu, Xiaoyi and Di, Sheng and Zhang, Zheng},
  journal={arXiv preprint arXiv:2603.00357},
  year={2026}
}

@article{sergeev2018horovod,
  title={Horovod: fast and easy distributed deep learning in TensorFlow},
  author={Sergeev, Alexander and Del Balso, Mike},
  journal={arXiv preprint arXiv:1802.05799},
  year={2018}
}

@article{wang2025reliable,
  title={Reliable and Resilient Collective Communication Library for LLM Training and Serving},
  author={Wang, Wei and Yu, Nengneng and Xiong, Sixian and Liu, Zaoxing},
  journal={arXiv preprint arXiv:2512.25059},
  year={2025}
}

@article{shoeybi2019megatron,
  title={Megatron-lm: Training multi-billion parameter language models using model parallelism},
  author={Shoeybi, Mohammad and Patwary, Mostofa and Puri, Raul and LeGresley, Patrick and Casper, Jared and Catanzaro, Bryan},
  journal={arXiv preprint arXiv:1909.08053},
  year={2019}
}

@article{zhao2023pytorch,
  title={Pytorch fsdp: experiences on scaling fully sharded data parallel},
  author={Zhao, Yanli and Gu, Andrew and Varma, Rohan and Luo, Liang and Huang, Chien-Chin and Xu, Min and Wright, Less and Shojanazeri, Hamid and Ott, Myle and Shleifer, Sam and others},
  journal={arXiv preprint arXiv:2304.11277},
  year={2023}
}

@article{zhang2024does,
  title={How Does Critical Batch Size Scale in Pre-training?},
  author={Zhang, Hanlin and Morwani, Depen and Vyas, Nikhil and Wu, Jingfeng and Zou, Difan and Ghai, Udaya and Foster, Dean and Kakade, Sham},
  journal={arXiv preprint arXiv:2410.21676},
  year={2024}
}

@article{ulfm,
  title={Fault tolerance of MPI applications in exascale systems: The ULFM solution},
  author={Losada, Nuria and Gonz{\'a}lez, Patricia and Mart{\'\i}n, Mar{\'\i}a J and Bosilca, George and Bouteiller, Aur{\'e}lien and Teranishi, Keita},
  journal={Future Generation Computer Systems},
  volume={106},
  pages={467--481},
  year={2020},
  publisher={Elsevier}
}

@inproceedings{ulfm-mpich,
  title={Lessons learned implementing user-level failure mitigation in mpich},
  author={Bland, Wesley and Lu, Huiwei and Seo, Sangmin and Balaji, Pavan},
  booktitle={2015 15th IEEE/ACM international symposium on cluster, cloud and grid computing},
  pages={1123--1126},
  year={2015},
  organization={IEEE}
}

@inproceedings{mohan2021checkfreq,
  title={{CheckFreq}: Frequent, {Fine-Grained} {DNN} checkpointing},
  author={Mohan, Jayashree and Phanishayee, Amar and Chidambaram, Vijay},
  booktitle={19th USENIX Conference on File and Storage Technologies (FAST 21)},
  pages={203--216},
  year={2021}
}

@inproceedings{eisenman2022check,
  title={{Check-N-Run}: A checkpointing system for training deep learning recommendation models},
  author={Eisenman, Assaf and Matam, Kiran Kumar and Ingram, Steven and Mudigere, Dheevatsa and Krishnamoorthi, Raghuraman and Nair, Krishnakumar and Smelyanskiy, Misha and Annavaram, Murali},
  booktitle={19th USENIX Symposium on Networked Systems Design and Implementation (NSDI 22)},
  pages={929--943},
  year={2022}
}

@inproceedings{maurya2024datastates,
  title={Datastates-llm: Lazy asynchronous checkpointing for large language models},
  author={Maurya, Avinash and Underwood, Robert and Rafique, M Mustafa and Cappello, Franck and Nicolae, Bogdan},
  booktitle={Proceedings of the 33rd international symposium on high-performance parallel and distributed computing},
  pages={227--239},
  year={2024}
}

@inproceedings{wang2023gemini,
  title={Gemini: Fast failure recovery in distributed training with in-memory checkpoints},
  author={Wang, Zhuang and Jia, Zhen and Zheng, Shuai and Zhang, Zhen and Fu, Xinwei and Ng, TS Eugene and Wang, Yida},
  booktitle={Proceedings of the 29th Symposium on Operating Systems Principles},
  pages={364--381},
  year={2023}
}

@inproceedings{nicolae2019veloc,
  title={Veloc: Towards high performance adaptive asynchronous checkpointing at large scale},
  author={Nicolae, Bogdan and Moody, Adam and Gonsiorowski, Elsa and Mohror, Kathryn and Cappello, Franck},
  booktitle={2019 IEEE International Parallel and Distributed Processing Symposium (IPDPS)},
  pages={911--920},
  year={2019},
  organization={IEEE}
}

@inproceedings{wan2024bytecheckpoint,
  title={{ByteCheckpoint}: A unified checkpointing system for large foundation model development},
  author={Wan, Borui and Han, Mingji and Sheng, Yiyao and Peng, Yanghua and Lin, Haibin and Zhang, Mofan and Lai, Zhichao and Yu, Menghan and Zhang, Junda and Song, Zuquan and others},
  booktitle={22nd USENIX Symposium on Networked Systems Design and Implementation (NSDI 25)},
  pages={559--578},
  year={2025}
}

@inproceedings{xiong2024superbench,
  title={{SuperBench}: Improving cloud {AI} infrastructure reliability with proactive validation},
  author={Xiong, Yifan and Jiang, Yuting and Yang, Ziyue and Qu, Lei and Zhao, Guoshuai and Liu, Shuguang and Zhong, Dong and Pinzur, Boris and Zhang, Jie and Wang, Yang and others},
  booktitle={2024 USENIX Annual Technical Conference (USENIX ATC 24)},
  pages={835--850},
  year={2024}
}

@inproceedings{jeon2019analysis,
  title={Analysis of {Large-Scale} {Multi-Tenant} {GPU} clusters for {DNN} training workloads},
  author={Jeon, Myeongjae and Venkataraman, Shivaram and Phanishayee, Amar and Qian, Junjie and Xiao, Wencong and Yang, Fan},
  booktitle={2019 USENIX Annual Technical Conference (USENIX ATC 19)},
  pages={947--960},
  year={2019}
}

@inproceedings{narayanan2021efficient,
  title={Efficient large-scale language model training on gpu clusters using megatron-lm},
  author={Narayanan, Deepak and Shoeybi, Mohammad and Casper, Jared and LeGresley, Patrick and Patwary, Mostofa and Korthikanti, Vijay and Vainbrand, Dmitri and Kashinkunti, Prethvi and Bernauer, Julie and Catanzaro, Bryan and others},
  booktitle={Proceedings of the international conference for high performance computing, networking, storage and analysis},
  pages={1--15},
  year={2021}
}

@inproceedings{rajbhandari2020zero,
  title={Zero: Memory optimizations toward training trillion parameter models},
  author={Rajbhandari, Samyam and Rasley, Jeff and Ruwase, Olatunji and He, Yuxiong},
  booktitle={SC20: international conference for high performance computing, networking, storage and analysis},
  pages={1--16},
  year={2020},
  organization={IEEE}
}

@inproceedings{jiang2024megascale,
  title={{MegaScale}: Scaling large language model training to more than 10,000 {GPUs}},
  author={Jiang, Ziheng and Lin, Haibin and Zhong, Yinmin and Huang, Qi and Chen, Yangrui and Zhang, Zhi and Peng, Yanghua and Li, Xiang and Xie, Cong and Nong, Shibiao and others},
  booktitle={21st USENIX Symposium on Networked Systems Design and Implementation (NSDI 24)},
  pages={745--760},
  year={2024}
}

@article{huang2019gpipe,
  title={Gpipe: Efficient training of giant neural networks using pipeline parallelism},
  author={Huang, Yanping and Cheng, Youlong and Bapna, Ankur and Firat, Orhan and Chen, Dehao and Chen, Mia and Lee, HyoukJoong and Ngiam, Jiquan and Le, Quoc V and Wu, Yonghui and others},
  journal={Advances in neural information processing systems},
  volume={32},
  year={2019}
}

@inproceedings{narayanan2019pipedream,
  title={PipeDream: Generalized pipeline parallelism for DNN training},
  author={Narayanan, Deepak and Harlap, Aaron and Phanishayee, Amar and Seshadri, Vivek and Devanur, Nikhil R and Ganger, Gregory R and Gibbons, Phillip B and Zaharia, Matei},
  booktitle={Proceedings of the 27th ACM symposium on operating systems principles},
  pages={1--15},
  year={2019}
}

@inproceedings{hu2024characterization,
  title={Characterization of large language model development in the datacenter},
  author={Hu, Qinghao and Ye, Zhisheng and Wang, Zerui and Wang, Guoteng and Zhang, Meng and Chen, Qiaoling and Sun, Peng and Lin, Dahua and Wang, Xiaolin and Luo, Yingwei and others},
  booktitle={21st USENIX Symposium on Networked Systems Design and Implementation (NSDI 24)},
  pages={709--729},
  year={2024}
}

@inproceedings{dash2024optimizing,
  title={Optimizing distributed training on frontier for large language models},
  author={Dash, Sajal and Lyngaas, Isaac R and Yin, Junqi and Wang, Xiao and Egele, Romain and Ellis, J Austin and Maiterth, Matthias and Cong, Guojing and Wang, Feiyi and Balaprakash, Prasanna},
  booktitle={ISC High Performance 2024 Research Paper Proceedings (39th International Conference)},
  pages={1--11},
  year={2024},
  organization={Prometeus GmbH}
}

@inproceedings{kokolis2025revisiting,
  title={Revisiting reliability in large-scale machine learning research clusters},
  author={Kokolis, Apostolos and Kuchnik, Michael and Hoffman, John and Kumar, Adithya and Malani, Parth and Ma, Faye and DeVito, Zachary and Sengupta, Shubho and Saladi, Kalyan and Wu, Carole-Jean},
  booktitle={2025 IEEE International Symposium on High Performance Computer Architecture (HPCA)},
  pages={1259--1274},
  year={2025},
  organization={IEEE}
}

@inproceedings{bouteiller2015plan,
  title={Plan b: Interruption of ongoing mpi operations to support failure recovery},
  author={Bouteiller, Aurelien and Bosilca, George and Dongarra, Jack J},
  booktitle={Proceedings of the 22nd European MPI Users' Group Meeting},
  pages={1--9},
  year={2015}
}

@article{bland2013post,
  title={Post-failure recovery of MPI communication capability: Design and rationale},
  author={Bland, Wesley and Bouteiller, Aurelien and Herault, Thomas and Bosilca, George and Dongarra, Jack},
  journal={The International Journal of High Performance Computing Applications},
  volume={27},
  number={3},
  pages={244--254},
  year={2013},
  publisher={Sage Publications Sage UK: London, England}
}

@inproceedings{laguna2014evaluating,
  title={Evaluating user-level fault tolerance for MPI applications},
  author={Laguna, Ignacio and Richards, David F and Gamblin, Todd and Schulz, Martin and de Supinski, Bronis R},
  booktitle={Proceedings of the 21st European MPI Users' Group Meeting},
  pages={57--62},
  year={2014}
}

@article{raffel2020c4,
  title={Exploring the limits of transfer learning with a unified text-to-text transformer},
  author={Raffel, Colin and Shazeer, Noam and Roberts, Adam and Lee, Katherine and Narang, Sharan and Matena, Michael and Zhou, Yanqi and Li, Wei and Liu, Peter J},
  journal={Journal of machine learning research},
  volume={21},
  number={140},
  pages={1--67},
  year={2020}
}

@article{llama3,
  title={The llama 3 herd of models},
  author={Grattafiori, Aaron and Dubey, Abhimanyu and Jauhri, Abhinav and Pandey, Abhinav and Kadian, Abhishek and Al-Dahle, Ahmad and Letman, Aiesha and Mathur, Akhil and Schelten, Alan and Vaughan, Alex and others},
  journal={arXiv preprint arXiv:2407.21783},
  year={2024}
}

@inproceedings{wan2025robust,
  title={Robust llm training infrastructure at bytedance},
  author={Wan, Borui and Liu, Gaohong and Song, Zuquan and Wang, Jun and Zhang, Yun and Sheng, Guangming and Wang, Shuguang and Wei, Houmin and Wang, Chenyuan and Lou, Weiqiang and others},
  booktitle={Proceedings of the ACM SIGOPS 31st Symposium on Operating Systems Principles},
  pages={186--203},
  year={2025}
}

@article{nanotron,
  title={The Ultra-Scale Playbook: Training LLMs on GPU Clusters. 2025},
  author={Tazi, Nouamane and Mom, Ferdinand and Zhao, Haojun and Nguyen, Phuc and Mekkouri, Mohamed and Werra, Leandro and Wolf, Thomas},
  journal={URl: https://huggingface. co/spaces/nanotron/ultrascaleplaybook},
  year={2025}
}

@article{maurya2026datastatesllm,
  title={DataStates-LLM: Scalable Checkpointing for Transformer Models Using Composable State Providers},
  author={Maurya, Avinash and Rafique, M Mustafa and Cappello, Franck and Nicolae, Bogdan},
  journal={arXiv preprint arXiv:2601.16956},
  year={2026}
}

@misc{1mgpuxai,
  author = {Matthew Gooding },
  title = {xAI targets one million GPUs for Colossus supercomputer in Memphis},
  url = {https://www.datacenterdynamics.com/en/news/xai-elon-musk-memphis-colossus-gpu/},
  year = {2024}
}

@misc{grok4,
  author = {xAI},
  title = {Grok 4},
  url = {https://x.ai/news/grok-4},
  year = {2025}
}

@misc{llama4100k,
  author = {Saif Hasan},
  title = {Scaling LLaMA4 Training to 100k},
  url = {https://atscaleconference.com/videos/scaling-llama4-training-to-100k/},
  year = {2026}
}

@article{paszke2019pytorch,
  title={Pytorch: An imperative style, high-performance deep learning library},
  author={Paszke, Adam and Gross, Sam and Massa, Francisco and Lerer, Adam and Bradbury, James and Chanan, Gregory and Killeen, Trevor and Lin, Zeming and Gimelshein, Natalia and Antiga, Luca and others},
  journal={Advances in neural information processing systems},
  volume={32},
  year={2019}
}

@article{kang2025elaswave,
  title={Elaswave: An elastic-native system for scalable hybrid-parallel training},
  author={Kang, Xueze and Xiang, Guangyu and Wang, Yuxin and Zhang, Hao and Fang, Yuchu and Zhou, Yuhang and Tang, Zhenheng and Lv, Youhui and Maman, Eliran and Wasserman, Mark and others},
  journal={arXiv preprint arXiv:2510.00606},
  year={2025}
}

@inproceedings{kumar2026reconfigurable,
  title={Reconfigurable Torus Fabrics for Multi-tenant ML},
  author={Vijaya Kumar, Abhishek and Ding, Eric and Devraj, Arjun and Bunandar, Darius and Singh, Rachee},
  booktitle={Proceedings of the 31st ACM International Conference on Architectural Support for Programming Languages and Operating Systems, Volume 2},
  pages={1547--1565},
  year={2026}
}
\bibliographystyle{abbrv}

\appendix

\section{Additional Evaluations and Details}
This appendix provides additional evaluation details and the results for \framework-HSDP that cannot be included in the main text due to the tight page limit.

\subsection{General Details of All \framework Evaluations}
\label{app:eval-details}

{\bf Testbed.}
All evaluations are conducted on up to 128 nodes of Polaris at the Argonne Leadership Computing Facility (ALCF). Each node is equipped with 4 NVIDIA 40GB A100 GPUs. Intra-node communication uses NVLink, and inter-node communication uses a high-speed Slingshot.

{\bf Software Stack.}
We use CUDA-aware OpenMPI 5.0.8~\footnote{https://docs.open-mpi.org/en/v5.0.x/.} as the communication backend for cross-replica gradient reduction. It is achieved by extending the open-source PyTorch~\footnote{We install a developer version PyTorch (2.9.0, forked in July 2025) from source with OpenMPI installation to enable MPI-aware PyTorch.} a customized process group backend (see \cref{app:impl-pgulfm}). For 3D parallelism, we build upon Nanotron~\cite{nanotron}, an open-source high-performance pre-training framework developed by Huggingface~\footnote{https://huggingface.co/nanotron.}. For HSDP, we directly use FSDP1+\texttt{HYBRID\_SHARD}~\cite{zhao2023pytorch} that PyTorch ships, and a self-implemented training loop that mirrors Nanotron's logic.

{\bf General Setup.}
For comparisons with failure-free references, we experiment on 128 nodes, i.e., 512 GPUs, to demonstrate \framework's scalability and robustness. Among these experiments, 256 GPUs are lost spread across the run. For comparisons with checkpoint-restart baselines, we experiment on 32 nodes, i.e., 128 GPUs, for faster turnaround time as 128-node jobs are very difficult to be queued. We remark that benchmarking against checkpoint approaches at a smaller scale actually favors baselines: the checkpoint-restart stall time (resource allocation, initialization, checkpoint save/load, first step cold start, rerun lost progress) increases dramatically as the system scales up, and is projected to be even longer than failure interval at hyper-scales~\cite{salpekar2026training, lee2026spare}, i.e., 10 min stall time for 100k system, but MTBF is projected to be $\sim5$ min for 600k system, while \framework does not suffer from these costs.

\textbf{Failure Injection.}
We use the failure simulator described in \cref{app:impl-fail-sim} plus NCCL watchdog timeout to abort the entire program in checkpoint-restart baselines, and to abort the faulty replicas in \framework. All failures are injected precisely during the gradient synchronization, which is the most challenging setting, especially for examining trajectory preservation. Every failure will result in partially reduced gradients, and \framework needs to correctly restore the affected ones else the curve might drift away.

For the single-failure runtime breakdown experiment, we assume a checkpoint interval of $N$, meaning that the baseline saves checkpoints every $N$ steps. We then inject a failure at step $1.5N$, corresponding to the expected midpoint of a checkpoint interval under an independent failure assumption. For example, when the checkpoint interval is 8, the failure is injected at step 12.

For the multi-failure experiment, we again use checkpoint interval $N$, and inject failures starting at step $1.5N$, then every $N$ steps thereafter, i.e., at $1.5N, 2.5N, \dots$.

\textbf{Profiling.}
For the single-failure experiment, in the baseline setting, we treat the first $N$ steps as warmup and exclude them from measurement. We start recording after step $N$, immediately before checkpoint save, denoted as timestamp $A$. The checkpoint save completion time is denoted as $B$. The job then runs for another $N/2$ steps, and the failure is injected at step $1.5N$, denoted as $C$. We record the time when the first run fully exits as $D$, which is also the start time of the recovery launch. The checkpoint load start and finish times are denoted as $E$ and $F$, respectively. The resumed training start time is denoted as $G$. After rerunning another $N/2$ steps, the completion time of step $1.5N$ is denoted as $H$.

The runtime breakdown is then computed as follows:
\begin{itemize}
    \item checkpoint save time = $B-A$
    \item normal run time before failure = $C-B$
    \item failure handling time = $D-C$
    \item checkpoint load time = $F-E$
    \item restart initialization time = $G-D-(F-E)$
    \item rerun time = $H-G$
\end{itemize}

For \framework-3D, we start recording at step $N$, then record the failure injection time, the time when recovery completes, and the time when step $1.5N$ completes.

For the multi-failure experiment, we measure the runtime between each pair of consecutive failures. For the GPU-hour based metric, we sample the current completed iteration at each GPU-hour boundary and use it to compute the total number of processed tokens.

\textbf{Metric Definitions.}
For the single-failure experiment, we report the \emph{raw} runtime breakdown that is defined by the runtime of each event described above. For the multi-failure experiment, we measure the runtime between consecutive failures.

Effective throughput is defined as:
\[
\text{Effective Throughput} = \frac{\text{Processed Tokens}}{\text{Runtime} \times \text{Alive GPUs}}.
\]

Under this definition, the baseline remains approximately flat across failures because it restores the same configuration after each restart and therefore makes the same progress per failure interval. In contrast, \framework operates with fewer and fewer surviving GPUs after successive failures.

For the processed-tokens versus GPU-hours metric, we record the completed iteration at each GPU-hour boundary and convert it to processed tokens. Together, these two metrics characterize resource efficiency and show that \framework can significantly improve GPU utilization under repeated failures.

\subsection{Additional Details of \framework-3D Experiments}

\textbf{Training Configuration.}
We train a LLaMA-style 7B model with micro-batch size $=1$, sequence length $=4096$, and BF16 training with gradient accumulation/synchronization in FP32. For the parallelization strategy, we use TP$=4$ within each node, PP$=2$. For comparisons with the failure-free reference, we use gradient accumulation $=128$ and DP$=64$; for comparisons with checkpoint-restart, we use accumulation $=8$ and DP$=16$. The choice of smaller gradient accumulation for the latter is for the wall-clock breakdown shown in \cref{fig:eval-cost-breakdown}, so that each component remains comparable. As we don't have access to an actual 100k system, this is due to practical consideration for showing every aspect of the checkpoint-restart strategy is non-trivial.

\subsection{Evaluations of \framework-HSDP}
\label{app:hsdp-eval}

\textbf{Training Configuration.}
We train a LLaMA-style 1B model with micro-batch size $=1$, sequence length $=4096$, gradient accumulation $=16$, HSDP shard size $=8$, and BF16 training with gradient accumulation/synchronization in FP32. For comparison with the failure-free reference, we have 64 replicas, while for the comparison with checkpoint-restart baseline, we have 16 replicas. The number of replica is deliberately chosen to remain the same as 3D parallelism's setting so we can simulate same number of replica failure. Consequently, the model size is reduced from 7B to 1B, as 7B would not fit in 8 40GB GPUs under HSDP, unless increasing the sharding size.

\begin{figure}[t]
\centering
\begin{subfigure}[t]{0.5\linewidth}
\centering
\includegraphics[width=\linewidth]{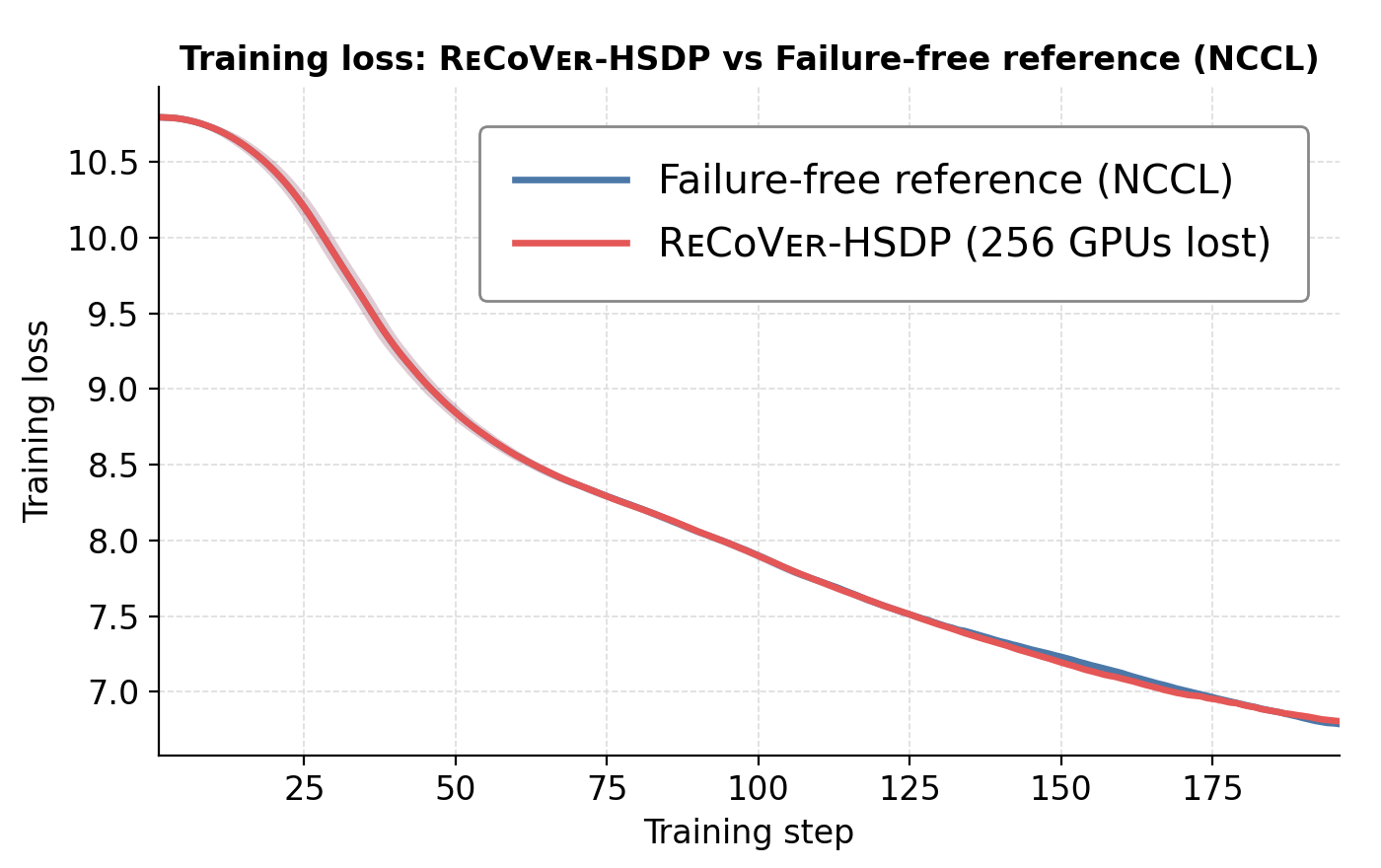}
\caption{}
\label{fig:eval-trajectory-loss-hsdp}
\end{subfigure}\hfill
\begin{subfigure}[t]{0.5\linewidth}
\centering
\includegraphics[width=\linewidth]{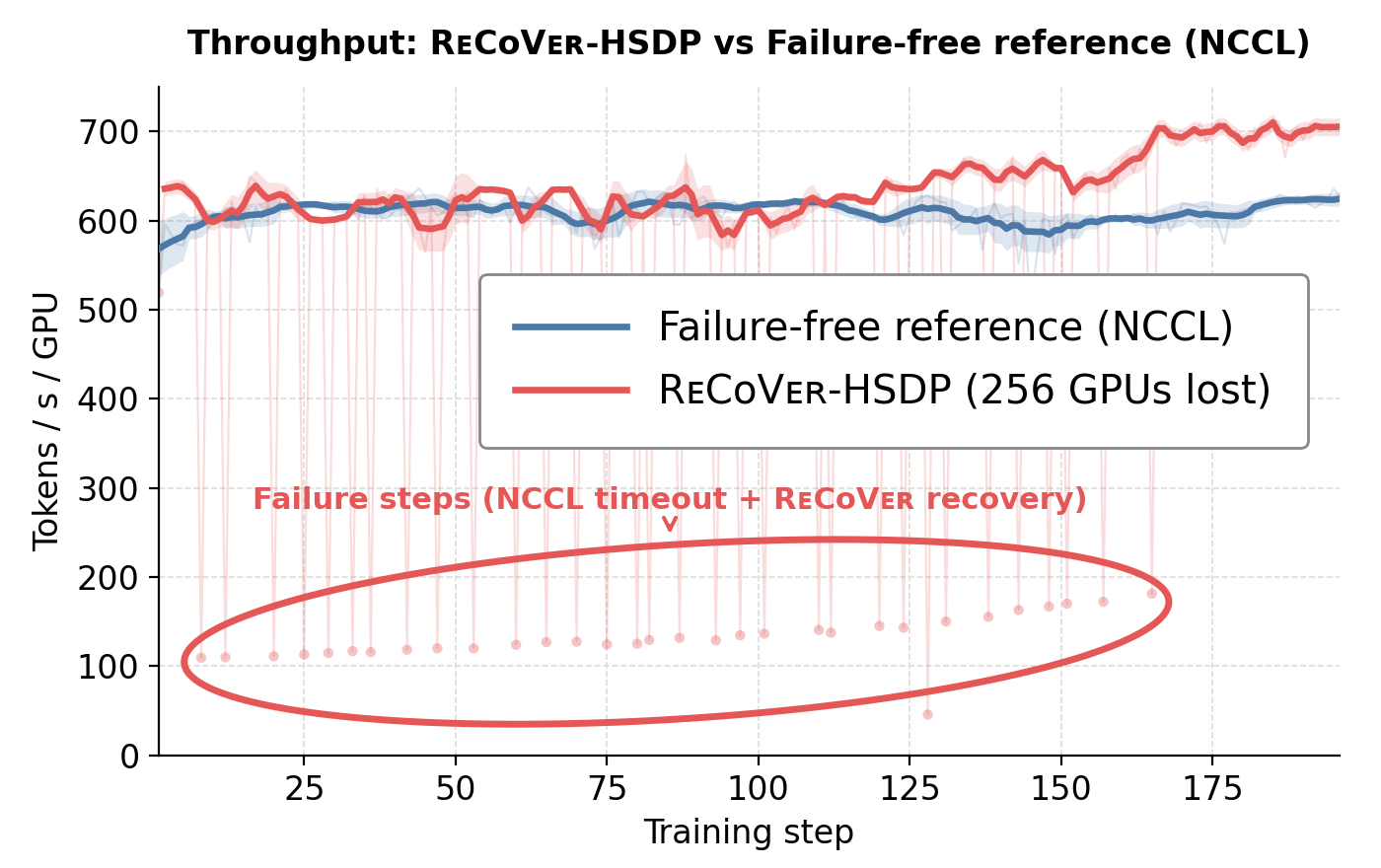}
\caption{}
\label{fig:eval-trajectory-throughput-hsdp}
\end{subfigure}
\caption{Trajectory preservation under $256$ GPU losses on a 512-GPU HSDP run. \subref{fig:eval-trajectory-loss}~The
\framework-HSDP training loss curve matches the failure-free
NCCL reference throughout the run, with no spikes or measurable deviation. \subref{fig:eval-trajectory-throughput}~The corresponding effective throughput: \framework-HSDP matches NCCL closely, and surpasses it after successive failures as \framework increases per-iteration workload and improves per-GPU utilization.}
\label{fig:eval-trajectory}
\end{figure}

\begin{figure}[t]
\centering
\begin{minipage}[c]{0.52\linewidth}
\centering
\begin{subfigure}[t]{\linewidth}
\centering
\vspace{10pt}
\includegraphics[width=\linewidth]{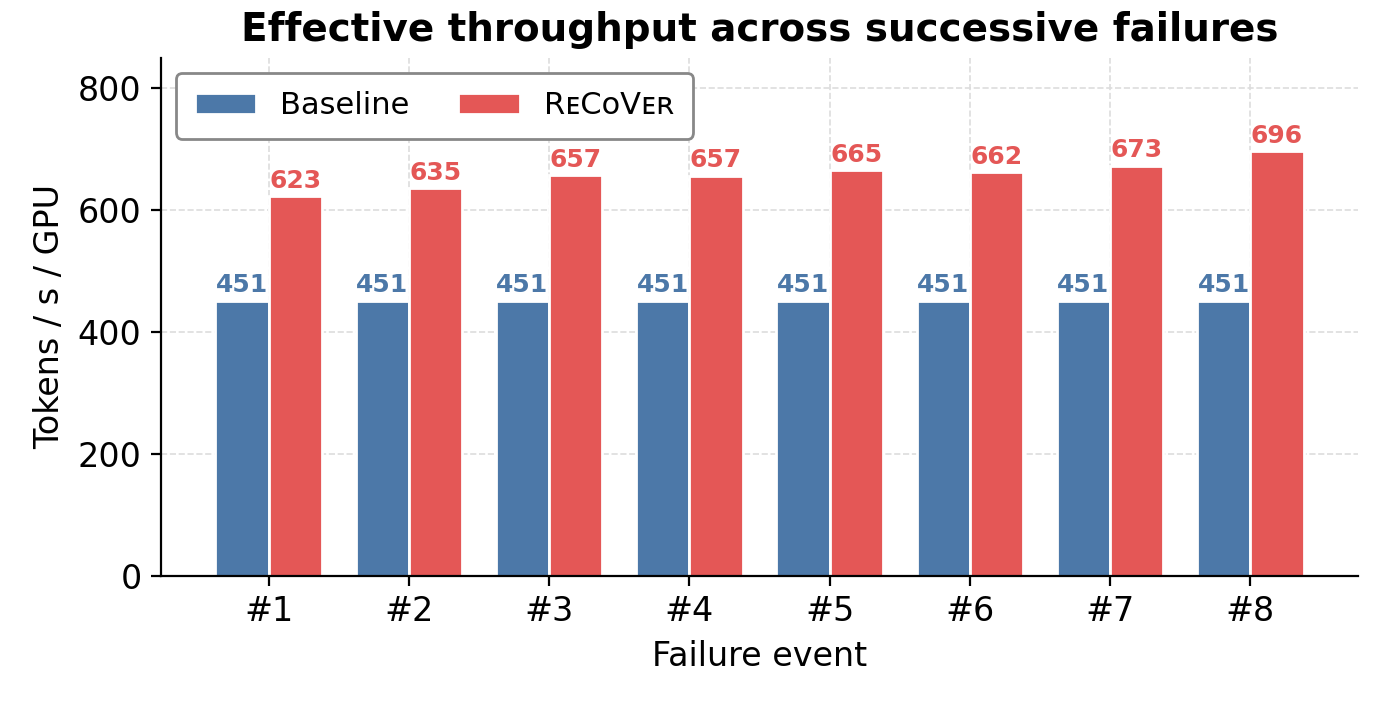}
\vspace{-20pt}
\caption{}
\label{fig:eval-cost-throughput-hsdp}
\end{subfigure}\\[4pt]
\begin{subfigure}[t]{\linewidth}
\centering
\vspace{2.5pt}
\includegraphics[width=\linewidth]{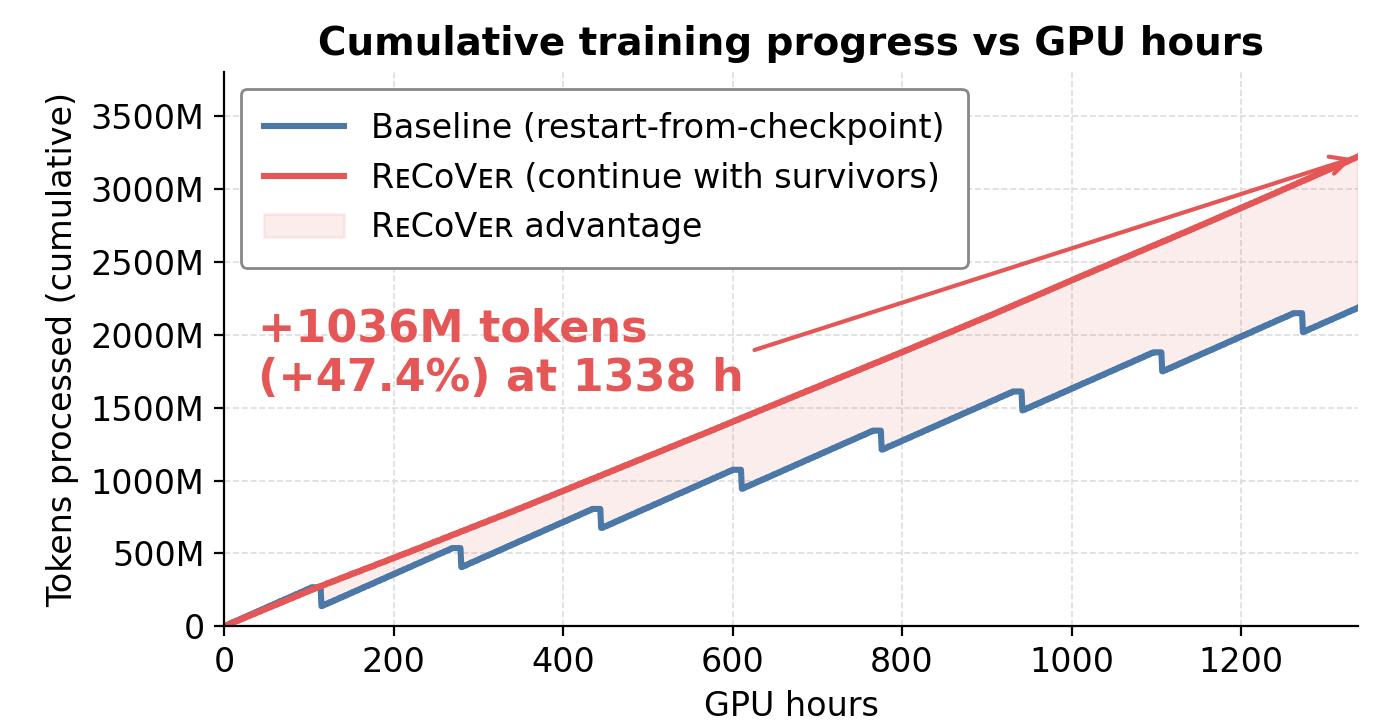}
\vspace{-12pt}
\caption{}
\label{fig:eval-cost-progress-hsdp}
\end{subfigure}
\end{minipage}\hfill
\begin{subfigure}[c]{0.48\linewidth}
\centering
\includegraphics[width=\linewidth]{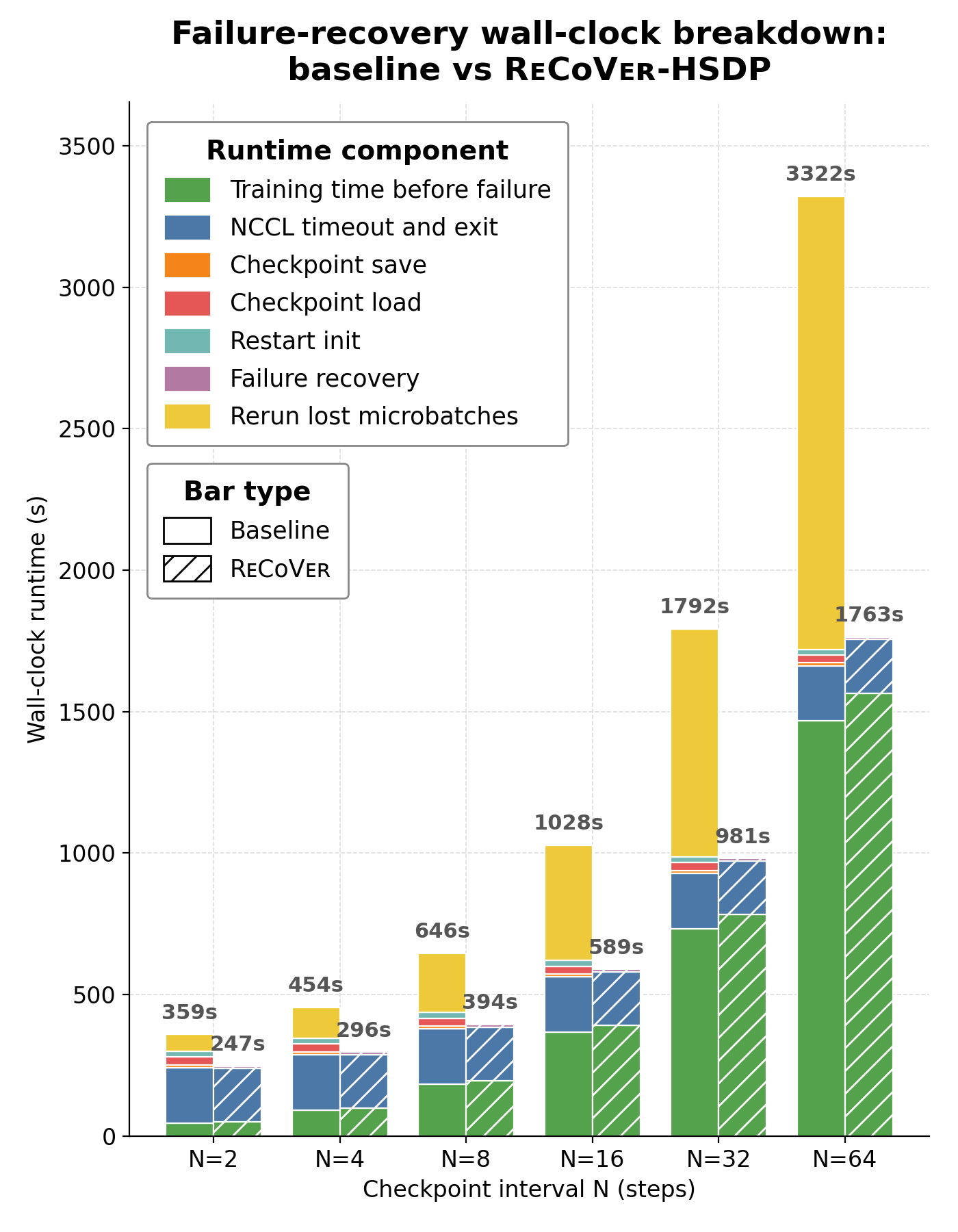}
\vspace{-20pt}
\caption{}
\label{fig:eval-cost-breakdown-hsdp}
\end{subfigure}
\caption{Cost comparison between \framework-HSDP and
restart-from-checkpoint. \subref{fig:eval-cost-throughput-hsdp}~Effective
throughput across successive failures; \framework-HSDP is consistently higher than checkpoint baseline. \subref{fig:eval-cost-progress-hsdp}~Cumulative training progress in tokens vs GPU-hours; \framework-HSDP processes $47.4\%$ more tokens at $1338$ GPU-hours.
\subref{fig:eval-cost-breakdown-hsdp}~Single-failure raw wall-clock breakdown swept over checkpoint interval $N$. \framework wins even at baseline's most favorable setting (i.e., $N=2$).}
\vspace{-10pt}
\label{fig:eval-cost-hsdp}
\end{figure}

\textbf{Evaluation Results.} Similar to the results of \framework-3D shown in \cref{sec:eval}, we present the comparison of \framework-HSDP against a failure-free reference and a standard checkpoint-restart baseline. In \cref{fig:eval-trajectory-loss-hsdp}, we see that \framework-HSDP again produces a loss curve that is indistinguishable to its failure-free run, and maintains an effective throughput that matches the NCCL reference and increases as failures keep happening. These results are mostly identical to \framework-3D case, except \framework-HSDP is not even slower at the beginning, possibly due to that each local shard of the model is smaller in 1B case (as we deliberately chose replica size to be the same, i.e., 8 GPUs a replica, so each shard is effectively smaller) so that OpenMPI's backend overhead is less pronounced. In \cref{fig:eval-cost-breakdown-hsdp}, we see the gap between checkpoint baseline and \framework-HSDP remains the same ratio as the ones in \framework-3D, except here the checkpoint-related overhead is less comparable to other components as that each HSDP iteration is longer than it was in 3D parallelism's case. And similarly, \ref{fig:eval-cost-throughput-hsdp} and \cref{fig:eval-cost-progress-hsdp} show similar benefit over checkpoint baselines. The gap is slightly smaller than in \framework-3D, which is reasonable as our improvement on resource utilization comes from pushing the hardware toward compute-bound regime. In 3D parallelism, communications are highly structural: high-frequency and high-volumes ones, such as the TP all-reduce is restrained to intra-node, while inter-node communication only handles PP send/recv and gradient all-reduce, which are much lighter than HSDP's frequent all-gather/reduce-scatter, that happens on every parameter's every forward/backward across all devices in each replica, which span across multiple nodes. On a system that is not deliberately designed/optimized for inter-node GPU communication, HSDP is much less scalable and efficient than 3D parallelism, which can be reflected by comparing \cref{fig:eval-cost-throughput} and \cref{fig:eval-cost-throughput-hsdp}: despite training a smaller model (1B), HSDP's effective throughput is only $\sim1/3$ of 3D parallelism's. Therefore, we remark that this gap difference is due to HSDP's inner structure and its interaction with our testbed hardware, not caused by \framework's design or implementation.

\section{Extended Terminology}
\label{app:extended-terms}

\cref{tab:extended-terms} consolidates the terms used throughout the paper.

\begin{table}[ht]
\centering
\small
\begin{tabularx}{\linewidth}{@{}lX@{}}
\toprule
Term & Description \\
\midrule
Microbatch                & one forward-backward pass done by a replica. \\
Gradient synchronization & gradient all-reduce across the replicas \\
Gradient accumulation     & gradient aggregation of
$G$ microbatches on each replica before gradient synchronization. \\
Iteration                 & the full cycle
of gradient accumulation, synchronization, model update; model parameters are updated exactly once per iteration. \\
Replica                   & One complete copy of the model. \\
Microbatch size   & Number of tokens a replica processes in one microbatch. \\
Global batch $B$          & Number of microbatches contributed to an iteration, summed across replicas. \\
$W_{\mathrm{init}}, G_{\mathrm{init}}$ & Initial replica count and initial grad-accum. \\
Major / minor     & Replica roles assigned by versatile-workload policy; see \cref{sec:versatile}. \\
Major-spare / minor-spare & Replica roles assigned by versatile-workload policy; see \cref{sec:versatile}. \\
Global rank               & A rank's index in the world communicator. \\
DP rank                   & A replica's index along the data-parallel dimension. \\
TP rank, PP rank          & A rank's index inside its replica's tensor / pipeline group. \\
MP rank                   & A rank's index within its replica (across TP and PP together). \\
Shard rank                & A rank's index inside its HSDP shard group. \\
Policy boundary           & The moment where a failure has exhausted spares and extra microbatches must be produced at the current iteration, and a new grad-accum must be adopted for the future iterations. \\
Policy boundary step      & The step to perform the extra microbatches when reaching policy boundary. \\
\bottomrule
\end{tabularx}
\caption{Consolidated terminology used in the main text and appendices.}
\label{tab:extended-terms}
\end{table}

\section{Implementation}
\label{app:impl}

This appendix describes how \framework's three protocol layers
(\cref{sec:framework}) are realized in open-source PyTorch~\cite{paszke2019pytorch}, and Nanotron~\cite{nanotron}. The implementation is laid out as a PyTorch-native ULFM communication backend (\texttt{ProcessGroupULFM}, written in C++) plus a Python control plane (\texttt{ulfm\_collectives/}) that turns the fault-tolerant
collective into a fault-tolerant optimizer step. We integrate the control plane with three pre-training stacks --- a flat DDP loop, a HSDP (FSDP1~\cite{zhao2023pytorch} with \texttt{HYBRID\_SHARD}) loop, and a Nanotron 3D(+EP) loop.

\paragraph{\texttt{ProcessGroupULFM}: the fault-tolerant backend.}\label{app:impl-pgulfm}
\texttt{ProcessGroupULFM} subclasses PyTorch's \texttt{ProcessGroup}
and is registered under the backend name \texttt{"ulfm"}; \framework
selects this backend for the cross-replica process group
$PG_{\mathrm{cross}}$ (the data-parallel group in 3D parallelism;
the FSDP \emph{replicate} group in HSDP). The backend exposes
exactly the four ULFM phases of \cref{sec:ulfm-guarded} as two
PyTorch-native collectives: \ulfmallreduce wraps an in-place sum
all-reduce with the four-phase \emph{detect--repair--record--reduce}
sequence, and \ulfmconsensus wraps the first three phases as a
barrier-like collective for failure view synchronization. Internally, both collectives drive the standard ULFM
primitives \texttt{MPIX\_Comm\_agree}, \texttt{MPIX\_Comm\_revoke},
and \texttt{MPIX\_Comm\_shrink} on the underlying MPI communicator,
and on every successful repair bump a monotone integer
\texttt{worldEpoch()} that the Python control plane stamps onto each
gradient-bucket snapshot to drive the world-epoch classification of
\cref{alg:gradient-restoration}. Each call returns a \texttt{WorkULFM} future
that carries failure metadata --- \texttt{has\_failures()},
\texttt{get\_failed\_ranks()}, the survivor census, and (when
versatile-workload tracking is enabled) per-role failure counts and
contribution counters --- so the Python hook can inspect a single
work object after \texttt{.wait()} and drive all downstream policy
logic from it. A pg-level \texttt{set\_quiesce(true/false)} latch
lets the control plane short-circuit any further bucket all-reduces to a pre-resolved future once a failure has been observed in the window, avoiding doing meaningless work that are meant to be restored even succeeded.

The same backend also carries the versatile-workload state for the
top protocol layer: per-rank atomic flags (\texttt{is\_minor\_}, \texttt{is\_spare\_}, \texttt{is\_boundary\_minor\_}) plus per-role atomic counters (\texttt{num\_major\_procs\_}, \texttt{num\_minor\_procs\_},
\texttt{num\_major\_spare\_procs\_}, \texttt{num\_minor\_spare\_procs\_},
\texttt{num\_boundary\_minor\_procs\_}) and per-rank contribution
counters (\texttt{contributed\_}, \texttt{target\_contribution\_},
plus a separate \texttt{boundary\_*} pair used during a policy boundary step). When a failure is detected, two helpers run \emph{within} the allreduce: \texttt{detect\_and\_recover\_failures} (i)~lists the failed ranks, (ii)~repairs the communicator and bumps the epoch, and \texttt{record\_and\_handling\_failure} (iii)~re-censuses survivors, (iv)~computes per-role
failure counts, (v)~evaluates the policy-boundary predicate
(major-failure with no major-spare \emph{or} minor-failure with no
minor-spare), (vi)~latches the
\texttt{atPolicyBoundary\_} flag if true, and (vii)~when no boundary
is crossed, runs \texttt{elect\_promotion} to atomically promote a
spare into the vacated role. Everything is then attached to the
\texttt{WorkULFM} that returns to Python, so $W_{\mathrm{cur}}$,
$C_{\mathrm{cur}}$, the per-role failure counts, and the boundary
verdict arrive at the policy layer in one collectively agreed
package.

\paragraph{\texttt{StepTxnOrchestrator}: per-iteration transaction state.}\label{app:impl-orch}
The orchestrator owns the iteration-local state: the list of
pre-reduce bucket snapshots
$\{(b, S(b), \epsilon(b), \mathrm{idx}(b))\}$, the set of buckets that have reduced cleanly under the current epoch, the deferred bucket queue that needs to be drained after backward computation finalizes, and the current
restoration plan. It exposes a single unified entry point,
\texttt{handle\_work\_completion(work, bucket\_index)}, that every future of \texttt{WorkULFM} invokes upon \texttt{.wait()}: on success it marks the bucket reduced under the current epoch; on failure it (i)~unpacks \texttt{FailureStats} and \texttt{RankTypeCounts} from the \texttt{WorkULFM}, (ii)~packages them with its own progress tracker (current microbatch index, total in-window, world epoch, current
rank/size) into a \texttt{FailureEvent}, (iii)~consults the policy, (iv)~latches \texttt{at\_policy\_boundary} and the restore mode, and (v)~either quiesces the pg or proceeds based on the policy decision. Two restore implementations live on the orchestrator: \texttt{restore\_gradients\_non\_blocking()} schedules per-snapshot \texttt{view.copy\_(snap)} on a dedicated CUDA stream and records an event that the trainer waits on right before the next backward, so the rewind is overlapped with the boundary step's forward; \texttt{restore\_gradients\_blocking()} performs the rewind plus a re-issued \ulfmallreduce on every stale bucket synchronously before the optimizer step, with a guarded retry path for the rare case in which the re-reduction itself hits a failure: if the failure is a non-boundary-crossing failure, retry the re-reduction; if it reaches policy boundary, break from the loop and enters the policy boundary step logic. After the optimizer step commits, \texttt{after\_successful\_commit()} drives the policy advance: \texttt{policy.advance\_policy()} returns the new
$(n_{\mathrm{maj}}, n_{\mathrm{min}}, n_{\mathrm{ms}}, n_{\mathrm{mi}})$
layout (\cref{alg:policy-advancement}), which is then pushed into the C++
backend via \texttt{set\_major\_minor\_split\_with\_spares} and
\texttt{update\_rank\_type\_counts}, the
\texttt{at\_policy\_boundary} latch is cleared, and
\texttt{set\_target\_contribution} primes the per-role microbatch
quotas for the next iteration.

\paragraph{TrainingManager: the microbatch state machine.}
The TrainingManager is the only component that touches the PyTorch module and the optimizer. Across all three integrations it preserves the same per-microbatch state machine: at $m=0$ in the window it runs \texttt{on\_iteration\_start()} and zeros the gradient buffers; on each microbatch, if the current restore plan is \textsc{non-blocking} (an extended boundary pass) it kicks off the asynchronous rewind onto a dedicated CUDA stream so it overlaps with the upcoming forward; before the backward of the first extended-pass microbatch, it calls
\texttt{wait\_restore\_before\_backward()} to synchronize the rewind; on every microbatch it consults
\texttt{pg.should\_contribute()} and zeros the loss for that microbatch when the rank has already met its quota for the iteration;
after synchronization it normalizes the accumulated gradient by the target global batch
(\texttt{effective\_batch\_size = $W_{\mathrm{init}} \cdot
G_{\mathrm{init}}$}, constant under \texttt{StaticWorldPolicy}),
runs the optimizer step, and calls
\texttt{after\_successful\_commit()} so the policy can advance the
role layout (\cref{alg:policy-advancement}) for the next iteration. The three
training-stack integrations described below differ only in (a)
which comm hook is registered on the model and (b) which entity
drives the outer per-iteration recovery loop; the shared microbatch
mechanics, the orchestrator, and the policy machinery are
unchanged.

\paragraph{Path 1 --- DDP-only, immediate hook.}
The flat-DDP path is the simplest and the only one that submits
MPI work \emph{inside} backward. \texttt{ULFMTrainingManager} wraps
the model in standard PyTorch \texttt{DistributedDataParallel} on the ULFM
\texttt{WORLD} process group and registers
\texttt{create\_ulfm\_recovery\_hook} as the comm hook. Each
backward triggers the hook once per gradient bucket: the hook
snapshots the bucket via \texttt{orch.on\_bucket\_snapshot}, posts \ulfmallreduce on $PG_{\mathrm{cross}} = \texttt{WORLD}$, and on the future's \texttt{then(...)} callback routes the resulting \texttt{WorkULFM} through \texttt{orch.handle\_work\_completion}.
Because the bucket schedule is a flat list with no pipeline or shard collectives running concurrently on the same device,
blocking inside the hook does not interfere with any other collective on the same rank, so the immediate path is preferred: The hook awaits \ulfmallreduce inline, so any failure is reported on the exact bucket whose reduction it interrupted, and the orchestrator runs its restore-plan logic before \texttt{train\_step} returns — the trainer's outer loop simply calls \texttt{train\_step} again for the next microbatch, with no recovery-aware control flow of its own. This path is used in the DDP-only single-replica-per-rank configuration as a reference implementation of the protocol mechanics; it is not the path used in experiments of \cref{sec:eval}.

\paragraph{Path 2 --- 3D parallelism on Nanotron, fp32-deferred hook.}
\framework-3D extends Nanotron with a \texttt{ULFMParallelContext} that places the data-parallel group on the ULFM backend and TP/PP/EP on NCCL with extended async-error-handling timeouts.
\texttt{NanotronULFMTrainingManager} binds the orchestrator to the DP process group and, when Nanotron's
\texttt{accumulate\_grad\_in\_fp32} is enabled (the common case for LLM pre-training), registers
\texttt{create\_ulfm\_fp32\_deferred\_hook} on the inner DDP. The hook does three things in one place: (i)~accumulates the bucket's bf16 gradients into Nanotron's contiguous fp32 grad accumulator
(\texttt{\_contiguous\_fp32\_grad\_buffer}) unconditionally, so that local accumulation is preserved even if the cross-replica pg is quiesced; (ii)~derives the contiguous slice of the fp32 buffer covering this bucket's parameters (asserting gap-freeness, which holds under DDP's reverse-registration bucketing) and snapshots that slice for failure rollback; and (iii)~queues the same slice \emph{view} on the orchestrator's deferred-bucket queue and returns a pre-resolved future, so DDP's \texttt{finalize\_backward}
never blocks on ULFM logic, and therefore the following PP \texttt{send/recv} on activation gradients won't timeout and kill the ranks in healthy replicas. The actual cross-replica all-reduce lands later, when the trainer's outer loop calls
\texttt{fire\_deferred\_allreduces}; because the queued buffer is a view into the fp32 accumulator's storage, the reduction lands in place and no scatter-back into per-parameter grads is needed. The outer per-iteration recovery loop in \texttt{trainer\_ulfm.py} then barriers on \texttt{mp\_pg} (the union of TP+PP+EP group) as the replica-consistency gate to ensure faulty replica's members are killed cleanly, run a \ulfmconsensus on \texttt{dp\_pg} to ensure consistent post-failure view across intra-replica ranks in healthy replicas, and dispatch on the resulting restore mode --- \textsc{skip} commits the optimizer step, \textsc{non-blocking} starts the async rewind and re-enters the loop with the boundary-extension microbatch count, and \textsc{blocking} runs a synchronous rewind-and-re-reduce that optionally crosses into the non-blocking branch if the re-reduce itself trips a policy boundary.

\paragraph{Path 3 --- HSDP, deferred hook.}
\framework-HSDP wraps the model with FSDP1 in
\texttt{HYBRID\_SHARD} mode, with the intra-replica
\texttt{shard\_pg} on NCCL and the cross-replica
\texttt{replicate\_pg} on \framework's ULFM backend.
\texttt{HSDPULFMTrainingManager} binds the orchestrator to
\texttt{replicate\_pg} and registers
\texttt{create\_ulfm\_hsdp\_hook} on the FSDP-wrapped model.
Critically, the hook is also \emph{deferred}: when FSDP fires it on the sync microstep with the unit's
\texttt{flat\_param.\_saved\_grad\_shard}, the hook snapshots the shard accumulator, enqueues the buffer reference on the orchestrator's deferred-bucket queue, and returns a pre-resolved future. Similarly, the non-overlapping scheme is to prevent a cross-replica ULFM logic on \texttt{replicate\_pg} co-occurring with intra-replica
reduce-scatter and all-gather collectives on \texttt{shard\_pg}, where the NCCL watchdog on \texttt{shard\_pg} could time out on healthy replicas and terminated the very ranks supposed to drive recovery. By deferring the cross-replica MPI work to after the inner microbatch loop has fully returned, the intra-replica NCCL traffic on \texttt{shard\_pg} runs to completion on a clean schedule. To preserve the invariant of \cref{eq:invariant} under this change, \texttt{HSDPULFMTrainingManager} also strips FSDP's default replicate-axis grad post-divide so that the gradient divisor is owned solely by the manager's
\texttt{\_get\_grad\_div\_factor} and stays constant at
$W_{\mathrm{init}} \cdot G_{\mathrm{init}}$ regardless of which replicas survive. The outer per-iteration recovery loop is then driven by \texttt{main\_hsdp.py}: run the inner microbatch loop, drain the deferred queue via \texttt{fire\_cross\_replica\_allreduces} (which calls \texttt{txn.fire\_deferred\_allreduces}), barrier over \texttt{shard\_pg}'s ranks as the replica-consistency gate, run a
\ulfmconsensus on \texttt{replicate\_pg} to ensure consistent post-failure view across intra-shard ranks, and dispatch on the resulting restore mode, same as Path~2.


\paragraph{Policies.}\label{app:impl-policies}
The Python \texttt{FaultTolerancePolicy} hierarchy contains two
concrete classes. \texttt{StaticWorldPolicy}, the policy described
in \cref{sec:versatile} and used in all \framework experiments,
implements the in-iteration boundary handling of
\cref{alg:policy-adjustment} (the boundary-step computation that
picks $G_{\mathrm{ext}}$ as the smallest integer with
$C_{\mathrm{cur}} + W_{\mathrm{cur}} \cdot G_{\mathrm{ext}} \geq B$
and the corresponding boundary-minor split) and the post-boundary
steady-state advance of \cref{alg:policy-advancement}.
\texttt{AdaptiveWorldPolicy} is the strawman of
\cref{alg:adaptive}: on every failure it returns
\textsc{repair-and-continue} with a blocking restore and no
boundary handling, so the global batch shrinks with the world. We
keep it in the codebase so the same training stack and the same
collective backend can be re-used as an elasticity-only baseline,
which isolates the contribution of the versatile-workload layer.

\paragraph{Replica-consistency gate.}
All three paths use the gate introduced in \cref{sec:integration} to abort a replica as a unit when one or more of its intra-replica ranks have failed. Path~1 trivially satisfies the gate because every rank is its own replica. Path~2 barriers on \texttt{mp\_pg} (the union of TP/PP/EP). Path~3 barriers over \texttt{shard\_pg}'s ranks.

\paragraph{Failure simulator.}\label{app:impl-fail-sim}
For reproducible evaluation, the framework ships a deterministic
failure simulator that either generates or consumes a YAML schedule of
$(\text{step}, \text{replica id}, \text{local rank},
\text{location})$ entries. At each injection point (forward, backward, gradient synchronization, optimizer), every rank checks whether its scheduled location matches the current one; if it does, the rank issues
\texttt{os.kill(os.getpid(), SIGKILL)} to faithfully simulate a
crash failure. The schedule is a pure function of
$(\text{parallelism spec}, \text{seed}, \text{count}, \text{step
range}, \text{location weights})$, so every rank generates the same
schedule at startup without any cross-rank broadcast.

\section{Additional Algorithmic Details}
\label{app:algos}

This appendix expands the six black-box helpers called from the
main-text \cref{alg:stepflow-full}, in the order they fire during
an iteration:
\callulfmallreduce (\cref{alg:ulfm-allreduce}) on every gradient
bucket of the last microbatch;
\callulfmconsensus (\cref{alg:ulfm-consensus}) once after the
bucket loop, as the iteration's final cross-replica gate;
\callhandlefail (\cref{alg:handle-failure}) and
\callgradrestore (\cref{alg:gradient-restoration}) if any of the
ULFM collectives above flagged a failure;
\callpolicyadjust (\cref{alg:policy-adjustment}) immediately
after, to decide whether the iteration must run extra microbatches;
and finally \callpolicyadvance (\cref{alg:policy-advancement}) once
a boundary iteration has committed, to install the next iteration's
role layout. We give each helper at the level of \emph{what it does
and why}, and link each to its concrete implementation in
\cref{app:impl}.

\paragraph{Notation and shared state.}
We reuse the symbols of \cref{sec:framework}. $PG_{\mathrm{cross}}$
is the cross-replica process group on the ULFM-aware backend;
$PG_{\mathrm{intra}}$ is each replica's intra-replica group on
NCCL. Each $PG$ carries a monotone integer $\epsilon_{\mathrm{cur}}$
(its \emph{world epoch}) that the backend increments on every
successful repair. When the bucket loop snapshots a gradient
bucket $b$ into $S(b)$, it tags the snapshot with the
$\epsilon_{\mathrm{cur}}$ in force at the time; we call $b$
\emph{stale} if $\epsilon(b) < \epsilon_{\mathrm{cur}}$ ---
i.e.\ its most recent reduction (if any) was issued under a now-shrunk
membership and would carry the wrong weight if mixed with
current-epoch reductions in the iteration sum.
$W_{\mathrm{cur}}, G_{\mathrm{cur}}, B$ are the current replica
count, current grad-accum factor, and target global batch from
\cref{sec:versatile}; $C_{\mathrm{cur}} = \sum_r C_r(t)$ is the
running contribution count.

The \texttt{Work} object returned by \cref{alg:ulfm-allreduce} and
\cref{alg:ulfm-consensus} carries the reduction's result (when one
occurred) plus a \emph{failure record} with three fields that the
remaining helpers consume:
\begin{itemize}[leftmargin=*,itemsep=2pt,topsep=2pt]
\item \texttt{role\_counts} $= (n_{\mathrm{maj}}, n_{\mathrm{min}}, n_{\mathrm{ms}}, n_{\mathrm{mi}}, n_{\mathrm{bm}})$:
      the post-failure (and, if applicable, post-promotion) population of each replica role on $PG_{\mathrm{cross}}$ ---
      majors, minors, major-spares, minor-spares, and boundary-minors. The policy reads this both to detect
      whether a spare absorbed the failure (one of $n_{\mathrm{ms}}, n_{\mathrm{mi}}$ shrank by one
      relative to the pre-failure count) and to know how many survivors of each role remain.
\item \texttt{contrib} $= C_{\mathrm{cur}}$: the microbatches that survivors have
      \emph{already} finished in this iteration at the moment of failure. The policy
      uses it at a boundary to size the extension: $G_{\mathrm{ext}}$ is chosen as the smallest integer
      with $C_{\mathrm{cur}} + W_{\mathrm{cur}} \cdot G_{\mathrm{ext}} \geq B$, so without an honest
      $C_{\mathrm{cur}}$ the extension would either overshoot or undershoot $B$.
\item \texttt{at\_boundary}: \textbf{true} iff a major failed with no major-spare or a minor failed with
      no minor-spare --- equivalently, the failure could not be absorbed by a spare and the iteration
      must be extended.
\end{itemize}

\subsection{ULFM-guarded collectives}
\label{app:ulfm-collectives}

\Cref{alg:ulfm-allreduce} is a fault-aware sum all-reduce: it rarely
reduces under a failed membership, and it never crashes on a failed
rank. The four phases of \cref{sec:ulfm-guarded} divide labour
cleanly: \emph{Detect} is a cheap probe placed \emph{before} any
data motion, so a stale membership is caught before a reduction is
posted; \emph{Repair} shrinks $PG_{\mathrm{cross}}$ to the
survivors and bumps $\epsilon_{\mathrm{cur}}$ so downstream code
can tell what was reduced under what; \emph{Record} produces the
failure record described above with the guarantee that every
survivor walks away with the same record (one all-reduce on the
per-replica role flags pins down \texttt{role\_counts} and
\texttt{contrib}; \texttt{at\_boundary} is a deterministic local
function of \texttt{role\_counts} and is therefore agreed by
construction; spare promotion, when it fires, is a small collective
election so the post-promotion \texttt{role\_counts} is the same
on every survivor); and \emph{Reduce} is the actual data motion,
which runs only if Detect passed. The implementation lives in
\texttt{ProcessGroupULFM} (\cref{app:impl-pgulfm}); the body of
\emph{Record} corresponds directly to the C++ helper
\texttt{record\_and\_handling\_failure} described there.

\begin{algorithm}[ht]
\small
\caption{\ulfmallreduce$(t, PG)$: fault-aware sum all-reduce.}
\label{alg:ulfm-allreduce}
\begin{algorithmic}[1]
\REQUIRE Tensor $t$, fault-tolerant process group $PG$.
\ENSURE \texttt{Work} carrying the result and a failure record.
\STATE \textbf{if} $PG$ was quiesced by an earlier failure this iteration \textbf{then return} a no-op \texttt{Work}.
\STATE \emph{Detect:} probe $PG$ for any rank that has failed since the last call on $PG$.
\IF{a failure was detected}
  \STATE \emph{Repair:} revoke pending operations on $PG$, shrink $PG$ to the survivors, and increment $\epsilon_{\mathrm{cur}}(PG)$.
  \STATE \emph{Record (collective):}
         census the per-replica role flags via one all-reduce on $PG$ to obtain \texttt{role\_counts} and \texttt{contrib};
         set \texttt{at\_boundary} $\gets$ (a major died with no major-spare) \textbf{or} (a minor died with no minor-spare);
         if \textbf{not} \texttt{at\_boundary}, run a small election on $PG$ that promotes one major-/minor-spare into the vacated role and re-census so \texttt{role\_counts} reflects the new layout;
         attach the failure record to \texttt{Work}.
  \STATE \textbf{return} \texttt{Work} (no reduction performed).
\ENDIF
\STATE \emph{Reduce:} if this rank is a spare and \texttt{at\_boundary} is \textbf{false}, set $t \gets 0$;\ then sum-reduce $t$ in place over $PG$.
\STATE \textbf{return} \texttt{Work} marked successful.
\end{algorithmic}
\end{algorithm}

\Cref{alg:ulfm-consensus} is the same primitive without the data
motion of phase~4. \Cref{alg:stepflow-full} fires it \emph{once
per iteration}, after every per-bucket \callulfmallreduce has
returned and after the intra-replica barrier on $PG_{\mathrm{intra}}$.
It serves an important purpose at this position: it converts any
asymmetric failure outcome from the bucket loop --- one rank's
\callulfmallreduce returning failure while a peer's returned
success --- into a globally agreed verdict, so every survivor sees the same failures together.
The implementation shares the entry point of \callulfmallreduce in
\texttt{ProcessGroupULFM} (\cref{app:impl-pgulfm}).

\begin{algorithm}[ht]
\small
\caption{\ulfmconsensus$(PG)$: fault-aware barrier with no data motion.}
\label{alg:ulfm-consensus}
\begin{algorithmic}[1]
\REQUIRE Fault-tolerant process group $PG$.
\ENSURE \texttt{Work} carrying a failure record (or success).
\STATE Run \emph{Detect}, \emph{Repair}, and \emph{Record} from \cref{alg:ulfm-allreduce} on $PG$.
\STATE \textbf{return} \texttt{Work}.
\end{algorithmic}
\end{algorithm}

\subsection{Failure handling}
\label{app:handle-failure}

\Cref{alg:handle-failure} bridges the collective backend and the
policy: it consumes the failure record produced by Record
(\cref{alg:ulfm-allreduce}), asks the policy what to do
(\callpolicyadjust), and installs the answer as state visible to
the rest of this iteration --- a latched \texttt{restore\_mode}
that \callgradrestore will read, a quiesce on $PG_{\mathrm{cross}}$
that short-circuits any further bucket all-reduce in this
iteration to a no-op (those buckets will be rolled back anyway),
and the boundary-minor split that controls which replicas contribute fewer microbatches on the extended pass. The implementation lives in
\texttt{StepTxnOrchestrator} (\cref{app:impl-orch}) as the unified
\texttt{handle\_work\_completion} entry point that every ULFM collective
invokes after \texttt{.wait()}.

\begin{algorithm}[ht]
\small
\caption{\textsc{handle\_work\_failure}$(W)$.}
\label{alg:handle-failure}
\begin{algorithmic}[1]
\REQUIRE \texttt{Work} object $W$ with the failure record produced by Record (\cref{alg:ulfm-allreduce}).
\STATE Build a \texttt{FailureEvent} $e$ with $\langle W.\texttt{role\_counts}, W.\texttt{contrib}, W.\texttt{at\_boundary}, m, \epsilon_{\mathrm{cur}}\rangle$.
\STATE $d \gets \callpolicyadjust(e)$.
\IF{$e.\texttt{at\_boundary}$}
  \STATE install $d$'s boundary-minor split on $PG_{\mathrm{cross}}$ so each rank queries the correct workload on the subsequent policy boundary step.
  \STATE quiesce $PG_{\mathrm{cross}}$ \COMMENT{stale buckets will be rolled back}
\ENDIF
\STATE latch the iteration's restore mode $\gets d.\texttt{restore\_mode}$.
\STATE invalidate the per-epoch ``already reduced'' bookkeeping (the new epoch makes it stale).
\end{algorithmic}
\end{algorithm}

\subsection{Gradient restoration}
\label{app:gradient-restoration}

\Cref{alg:gradient-restoration} enforces the correctness invariant
of \cref{sec:instep}: every contribution admitted into the
iteration sum must come from the same membership of
$PG_{\mathrm{cross}}$, or the per-microbatch contributions carry
mismatched weights and the iteration gradient is corrupted. The
world-epoch tag makes the check local --- a bucket is stale iff
its tag predates the current epoch --- and a stale bucket must be
rewound from $S(b)$ before its content can be admitted under the
repaired membership.

The mode latched by \callpolicyadjust controls \emph{when} the
rewind fires. A non-boundary failure with a spare available leaves
the iteration's total microbatch count unchanged, so the rewind
plus a re-issued reduction must finish before the optimizer step
--- this is the blocking branch, which calls \callulfmallreduce
synchronously on each stale bucket. A boundary failure already
requires extra microbatches, so the rewind is scheduled on a side
CUDA stream and overlapped with the extended pass's forward; the
re-reduction then happens implicitly when the extended pass's all-reduce on the bucket fires on the repaired membership,
with no extra round-trip needed. The implementation
(\cref{app:impl-orch}) is the orchestrator's
\texttt{restore\_gradients\_blocking} and
\texttt{restore\_gradients\_non\_blocking} routines, dispatched
on the latched \texttt{restore\_mode}.

\begin{algorithm}[ht]
\small
\caption{\textsc{gradient\_restoration}.}
\label{alg:gradient-restoration}
\begin{algorithmic}[1]
\REQUIRE Bucket bookkeeping $\mathcal{B}$, current epoch $\epsilon_{\mathrm{cur}}$, latched restore mode.
\STATE \textbf{if} restore mode $=$ \textsc{skip} \textbf{then return}.
\STATE $\mathit{stale} \gets \{b \in \mathcal{B} : \epsilon(b) < \epsilon_{\mathrm{cur}}\}$.
\IF{restore mode $=$ \textsc{non-blocking}}
  \STATE \COMMENT{policy boundary: overlap rewind with the upcoming extended-pass forward.}
  \STATE on a dedicated CUDA stream, copy $S(b) \to b$ for every stale $b$;\ record an event for the next backward to wait on.
  \STATE clear $\mathcal{B}$ and unquiesce $PG_{\mathrm{cross}}$ \COMMENT{the extended pass will repopulate $\mathcal{B}$ with fresh snapshots and re-reduce on $\epsilon_{\mathrm{cur}}$ as it goes}
\ELSE
  \STATE \COMMENT{non-boundary: rewind \emph{and} re-reduce before the optimizer step.}
  \FORALL{stale bucket $b$}
    \STATE copy $S(b) \to b$;\ then $W \gets \callulfmallreduce (b, PG_{\mathrm{cross}})$ and wait.
    \STATE \textbf{if} $W$ failed \textbf{then return} \COMMENT{the caller cross-syncs replicas and re-enters \callhandlefail}
  \ENDFOR
  \STATE unquiesce $PG_{\mathrm{cross}}$;\ restore mode $\gets$ \textsc{skip}.
\ENDIF
\end{algorithmic}
\end{algorithm}

\subsection{Policy hooks}
\label{app:policy-hooks}

\Cref{alg:policy-adjustment} is the policy's response to a failure
inside an iteration, and is the only place where the loop bound
$P(\mathrm{major})$ in \cref{alg:stepflow-full} can change. The
two branches follow directly from whether a spare absorbed the
loss. If yes (the non-boundary case), the failure is invisible to
the iteration's totals: Record already promoted a spare in
\cref{alg:ulfm-allreduce}, and from this point on it carries the
vacated rank's microbatch quota. The policy returns with $P$
unchanged --- in particular $P(\mathrm{major})$ unchanged --- so
when control returns to \cref{alg:stepflow-full} the outer
\textbf{while} terminates as planned and the iteration commits
with $\sum_r C_r = B$ unchanged. If no (the boundary case), the
iteration must run extra microbatches to recover the lost
contributions: the policy picks the smallest $G_{\mathrm{ext}}$
with $C_{\mathrm{cur}} + W_{\mathrm{cur}} \cdot G_{\mathrm{ext}}
\geq B$, marks $W_{\mathrm{cur}} \cdot G_{\mathrm{ext}} - (B -
C_{\mathrm{cur}})$ survivors as \emph{boundary minors} that
contribute one fewer extra microbatch (so the total lands at
exactly $B$), and grows $P(\mathrm{major})$ by $G_{\mathrm{ext}}$
(boundary minors by $G_{\mathrm{ext}} - 1$). When the inner
bucket loop exits and the outer \textbf{while} re-tests
$m < P(\mathrm{major})$, the test is now true and the iteration
runs the extended pass. The implementation
(\cref{app:impl-policies}) is \texttt{StaticWorldPolicy.on\_failure},
with the boundary branch in \texttt{\_on\_policy\_boundary}.

\begin{algorithm}[ht]
\small
\caption{\textsc{policy\_adjustment}$(e)$: in-iteration response (may grow $P(\mathrm{major})$).}
\label{alg:policy-adjustment}
\begin{algorithmic}[1]
\REQUIRE A \texttt{FailureEvent} $e$ produced by \callhandlefail.
\ENSURE Updated $P$, $\rho$, plus a \texttt{PolicyDecision} carrying \texttt{restore\_mode}.
\IF{\textbf{not} $e.\texttt{at\_boundary}$}
  \STATE \COMMENT{a spare in the failed role was already promoted in Record; $P(\mathrm{major})$ stays the same.}
  \STATE refresh the policy's tracked counts ($W_{\mathrm{cur}}$, spares) from $e.\texttt{role\_counts}$.
  \STATE \textbf{return} $P$, $\rho$ unchanged (except the promoted spare); \texttt{restore\_mode} $=$ \textsc{blocking}.
\ELSE
  \STATE \COMMENT{spares of the failed role are exhausted; \emph{grow} $P(\mathrm{major})$ to extend the iteration.}
  \STATE $G_{\mathrm{ext}} \gets$ smallest integer $\geq 1$ with $C_{\mathrm{cur}} + W_{\mathrm{cur}} \cdot G_{\mathrm{ext}} \geq B$.
  \STATE $n_{\mathrm{bdry}} \gets W_{\mathrm{cur}} \cdot G_{\mathrm{ext}} - (B - C_{\mathrm{cur}})$ \COMMENT{boundary-minor count; $0$ if $B$ divides exactly}
  \STATE $P(\mathrm{major}) \gets P(\mathrm{major}) + G_{\mathrm{ext}}$;\ \ $P(\text{boundary-minor}) \gets P(\mathrm{major}) - 1$;\ \ designate $n_{\mathrm{bdry}}$ survivors as boundary minors and refresh $\rho$ on each rank.
  \STATE \textbf{return} updated $P$, $\rho$;\ \texttt{restore\_mode} $=$ \textsc{non-blocking}.
\ENDIF
\end{algorithmic}
\end{algorithm}

\Cref{alg:policy-advancement} fires only once per policy-boundary
iteration, after the optimizer step has committed: it installs the
steady-state role layout the surviving world will run from the
\emph{next} iteration onwards. The choice of $G_{\mathrm{cur}}$ is
the smallest factor that lets $W_{\mathrm{cur}}$ survivors cover
$B$ with a single integer grad-accum; the residue $R_{\mathrm{cur}}
= B - n_{\mathrm{maj}} \cdot G_{\mathrm{cur}}$ is absorbed by at
most one minor. Any extra survivors are reserved as spares (mostly
major-spares, with at least one minor-spare reserved when a minor
exists) so the next failure has a chance of landing in the
non-boundary branch of \callpolicyadjust. The implementation
(\cref{app:impl-policies}) is \texttt{StaticWorldPolicy.advance\_policy},
which the orchestrator drives from \texttt{after\_successful\_commit}
and pushes into the C++ backend via
\texttt{set\_major\_minor\_split\_with\_spares}.

\begin{algorithm}[ht]
\small
\caption{\textsc{policy\_advancement}: post-boundary steady-state.}
\label{alg:policy-advancement}
\begin{algorithmic}[1]
\REQUIRE Survivor count $W_{\mathrm{cur}}$, target $B$.
\ENSURE New role layout $(G_{\mathrm{cur}}, n_{\mathrm{maj}}, n_{\mathrm{min}}, R_{\mathrm{cur}}, n_{\mathrm{ms}}, n_{\mathrm{mi}})$ pushed to $PG_{\mathrm{cross}}$.
\STATE $G_{\mathrm{cur}} \gets$ smallest integer with $W_{\mathrm{cur}} \cdot G_{\mathrm{cur}} \geq B$.
\STATE $n_{\mathrm{maj}} \gets \lfloor B / G_{\mathrm{cur}} \rfloor$;\ \ $R_{\mathrm{cur}} \gets B - n_{\mathrm{maj}} \cdot G_{\mathrm{cur}}$;\ \ $n_{\mathrm{min}} \gets 1$ if $R_{\mathrm{cur}} > 0$ else $0$.
\STATE Allocate the remaining $W_{\mathrm{cur}} - n_{\mathrm{maj}} - n_{\mathrm{min}}$ ranks as major-spares and minor-spares; reserve one minor-spare when $n_{\mathrm{min}} = 1$ and $W_{\mathrm{cur}} - n_{\mathrm{maj}} - n_{\mathrm{min}} \geq 2$, otherwise all major-spares.
\STATE Push the new layout to $PG_{\mathrm{cross}}$, clear the policy-boundary flag, and update each surviving rank's role $\rho$ accordingly.
\STATE \textbf{return} the new policy $P$ and refreshed role $\rho$ for this rank.
\end{algorithmic}
\end{algorithm}

\subsection{\textsc{AdaptiveWorldPolicy}: a strawman repair-and-continue baseline}
\label{app:adaptive}

\Cref{alg:adaptive} is a drop-in replacement for
\cref{alg:policy-adjustment} that does the bare minimum: on any
failure it returns \textsc{blocking} restore with
\texttt{at\_boundary = false}, regardless of whether spares
exist; \callpolicyadvance is never called because no boundary is
ever crossed. The iteration commits with $W_{\mathrm{cur}} \cdot
G_{\mathrm{cur}} < B$ --- the global batch shrinks with the world
and the trajectory drifts. The implementation
(\cref{app:impl-policies}) is \texttt{AdaptiveWorldPolicy} on the
same orchestrator and ULFM backend as the static policy. We keep
it as an isolation baseline: pairing the same ULFM-guarded
collective and in-step rewind with this minimal policy reproduces
what prior keep-alive
frameworks~\cite{li2023elastic,salpekar2026training} provide, and
shows what versatile workload contributes on top.

\begin{algorithm}[ht]
\small
\caption{\textsc{AdaptiveWorldPolicy} (strawman).}
\label{alg:adaptive}
\begin{algorithmic}[1]
\REQUIRE A \texttt{FailureEvent} $e$.
\STATE \textbf{return} \texttt{PolicyDecision}(restore\_mode = \textsc{blocking}, at\_boundary = \textbf{false}). \COMMENT{$PG_{\mathrm{cross}}$ was repaired in phase~2 of \cref{alg:ulfm-allreduce}; the iteration commits with effective batch $W_{\mathrm{cur}} \cdot G_{\mathrm{cur}} < B$.}
\end{algorithmic}
\end{algorithm}

\section{Numerical Walk-through of a Boundary Step}
\label{app:walkthrough}

This appendix traces the $W_{\mathrm{init}} = 32$,
$G_{\mathrm{init}} = 8$ example of \cref{fig:workload} (so $B =
256$), making explicit how the contribution count $C_r(t)$ on each
replica accumulates and how the world-epoch tagging keeps every
contribution under the same membership of $PG_{\mathrm{cross}}$.
Recall the iteration schedule from \cref{alg:stepflow-full}: every
replica runs all $P(\mathrm{major})$ microbatches with local
accumulation, and only at the last microbatch does the bucket loop
fire per-bucket \ulfmallreduce on $PG_{\mathrm{cross}}$, followed
by a single barrier on $PG_{\mathrm{intra}}$ and a single
\ulfmconsensus on $PG_{\mathrm{cross}}$.

Let $\mathcal{R} = \{r_1, \dots, r_{32}\}$ denote the initial
replicas. Let $\epsilon_0$ be the world epoch on
$PG_{\mathrm{cross}}$ before the failure, and write
$\epsilon_1 = \epsilon_0 + 1$ for the post-repair epoch
that the Repair phase of \callulfmallreduce installs. Suppose
that at iteration $t^\star$ replica $r_{32}$ fails partway through
the bucket loop --- specifically, during the all-reduce of some
bucket $b^\star$.

\paragraph{Pre-failure iterations.}
For every iteration $t < t^\star$, every replica is a major with
$P(\mathrm{major}) = G_{\mathrm{init}} = 8$, so $C_r(t) = 8$ for
all $r$. The bucket loop reduces every bucket cleanly under epoch
$\epsilon_0$ and the iteration commits with
$\sum_{r=1}^{32} C_r(t) = 32 \cdot 8 = 256 = B$, satisfying
\cref{eq:invariant}.

\paragraph{State at the moment of failure (iteration $t^\star$).}
By the time the bucket loop is running, every replica has already
completed all $8$ microbatches of forward+backward and has each
microbatch's contribution recorded against its
\texttt{contrib} counter on $PG_{\mathrm{cross}}$. The Detect probe
of \callulfmallreduce on $b^\star$ surfaces $r_{32}$'s failure;
Repair shrinks $PG_{\mathrm{cross}}$ to the $31$ survivors and
bumps the world epoch to $\epsilon_1=\epsilon_0+1$; Record runs the
role-flag all-reduce among survivors and reports
$\texttt{contrib} = C_{\mathrm{cur}} = 31 \cdot 8 = 248$
(the contributions $r_{32}$ had logged are dropped because $r_{32}$
is no longer in the membership over which the census all-reduce
runs). \texttt{at\_boundary} is set to \textbf{true}: $r_{32}$ was
a major and no major-spare exists at iteration $t^\star$ to absorb
the loss.
At this moment, the survivors hold buckets in three positions
relative to $b^\star$:
\begin{itemize}
\item \emph{Buckets that completed before $b^\star$ (epoch
    $\epsilon_0 < \epsilon_1$).} These reduced cleanly under the
    full $32$-replica membership and would now carry the wrong
    weight if mixed with $31$-replica reductions in the same sum.
    They are stale per \cref{alg:gradient-restoration}: their
    pre-reduce snapshots $S(b)$ are scheduled to be copied back
    into the bucket buffers and will be re-reduced at $\epsilon_1$.
\item \emph{Bucket $b^\star$ itself.} Its all-reduce returned the
    failure signal, so its post-reduce content is undefined; it is
    rewound from $S(b^\star)$ and will be re-reduced at $\epsilon_1$.
\item \emph{Buckets that hadn't been visited yet.} Their snapshots
    were never taken, and their contents are still locally
    accumulated partial sums; they will get fresh snapshots
    tagged $\epsilon_1$ when the bucket loop resumes after
    recovery.
\end{itemize}

\paragraph{The boundary extension.}
\callpolicyadjust receives $C_{\mathrm{cur}} = 248$,
$W_{\mathrm{cur}} = 31$ and \texttt{at\_boundary = true}, and
chooses the smallest $G_{\mathrm{ext}}$ with $C_{\mathrm{cur}} +
W_{\mathrm{cur}} \cdot G_{\mathrm{ext}} \geq B$:
\[
G_{\mathrm{ext}} = \left\lceil \frac{B - C_{\mathrm{cur}}}{W_{\mathrm{cur}}} \right\rceil
= \left\lceil \frac{256 - 248}{31} \right\rceil = 1.
\]
The overshoot is $C_{\mathrm{cur}} + W_{\mathrm{cur}} \cdot
G_{\mathrm{ext}} - B = 248 + 31 - 256 = 23$, so $23$ of the $31$
survivors are reassigned from major to \emph{boundary minor} ---
the role that contributes $G_{\mathrm{ext}} - 1 = 0$ extra
microbatches. The remaining $8$ survivors stay majors and
contribute the full $G_{\mathrm{ext}} = 1$ extra microbatch each.
Equivalently, $P(\mathrm{major})$ grows from $8$ to $9$ while
$P(\mathrm{boundary\text{-}minor})$ stays at $8$. \callgradrestore
is latched in non-blocking mode and schedules the rewinds on a
side CUDA stream; the bucket bookkeeping $\mathcal{B}$ is cleared.

\paragraph{The extended pass.}
Control returns to \cref{alg:stepflow-full}'s outer
\textbf{while}, which now re-tests $m < P(\mathrm{major})$ against
the new $P(\mathrm{major}) = 9$ and re-enters with $m=9$. Every
survivor runs forward and backward on microbatch $9$. The $8$
majors locally accumulate (their $P(\rho)=9$); the $23$ boundary
minors zero this microbatch's gradient (their $P(\rho)=8$). After
backward, the bucket loop fires again on $PG_{\mathrm{cross}}$ at
epoch $\epsilon_1$, takes fresh
snapshots for every bucket, and reduces each one cleanly --- there
is no extra round of re-reduction needed, because the rewinds of
stale buckets have already overlapped with this microbatch's
forward and the buckets now carry the union of all $9$
microbatches' (for majors, boundary-minors still have 8) worth of locally-accumulated contributions. The
post-loop barrier on $PG_{\mathrm{intra}}$ and \callulfmconsensus
on $PG_{\mathrm{cross}}$ pass cleanly, and the iteration commits
its optimizer step.

\paragraph{Verifying the invariant at iteration $t^\star$.}
Every bucket admitted into iteration $t^\star$'s gradient was
reduced exactly once under the $31$-replica membership at epoch
$\epsilon_1$. Counting per-replica contributions,
\[
\sum_{r \in \text{survivors}(t^\star)} C_r(t^\star)
= \underbrace{31 \cdot 8}_{\text{31 survivors @ 8}}
\;+\; \underbrace{8 \cdot 1}_{\substack{\text{8 majors}\\ \text{$\times$ 1 extra}}}
\;+\; \underbrace{23 \cdot 0}_{\substack{\text{23 bdry-min}\\ \text{$\times$ 0 extra}}}
= 248 + 8 + 0
= 256 = B,
\]
as \cref{eq:invariant} demands. The per-replica counts ---
$8$ replicas at $9$ microbatches and $23$ replicas at $8$
microbatches --- match panel~(ii) of \cref{fig:workload} exactly.
$r_{32}$'s data partition drops out of the run for good from
iteration $t^\star + 1$ onwards, while the $31$ survivors advance
through their own partitions slightly faster than the failure-free
schedule prescribed. Because the pre-training stream is effectively
infinite, this is statistically indistinguishable from having
sampled a different random shuffle of the same stream from the
outset; \cref{app:proof} formalizes the argument.

\paragraph{Restore-stream timing.}
The stale-bucket rewinds run on a CUDA stream disjoint from the
compute stream, kicked off as soon as
\callgradrestore latches the non-blocking mode. The rewind itself is a snapshot-to-buffer memcpy at typical
bucket sizes ($O(100\,\text{MB})$ of fp32). The compute stream reaches the first backward on the extra microbatches only after the rewind stream has finished.

\paragraph{Post-boundary steady state, iteration $t^\star + 1$.}
Once iteration $t^\star$ commits, \callpolicyadvance is invoked
with $W_{\mathrm{cur}} = 31$, $B = 256$ and previous
$G_{\mathrm{cur}} = 8$. It bumps $G_{\mathrm{cur}}$ until
$W_{\mathrm{cur}} \cdot G_{\mathrm{cur}} \geq B$, giving
$G_{\mathrm{cur}} = 9$. Then $n_{\mathrm{maj}} = \lfloor 256/9
\rfloor = 28$ and $R_{\mathrm{cur}} = 256 - 28 \cdot 9 = 4$, so
$n_{\mathrm{min}} = 1$. The remaining $W_{\mathrm{cur}} -
(n_{\mathrm{maj}} + n_{\mathrm{min}}) = 2$ ranks are spares ---
$1$ major-spare and $1$ minor-spare, since $n_{\mathrm{min}} = 1$ and $W_{\mathrm{cur}} -
(n_{\mathrm{maj}} + n_{\mathrm{min}}) > 1$. This new
layout is pushed to $PG_{\mathrm{cross}}$ and matches panel~(iii)
of \cref{fig:workload}. Verifying the invariant for the new
steady state,
\[
\sum_{r \in \text{survivors}(t^\star+1)} C_r(t^\star+1)
= \underbrace{28 \cdot 9}_{\text{majors}}
\;+\; \underbrace{1 \cdot 4}_{\text{minor at $R_{\mathrm{cur}}$}}
\;+\; \underbrace{1 \cdot 0 + 1 \cdot 0}_{\text{spares}}
= 252 + 4 + 0 = 256 = B.
\]
The two spares run their full forward and backward but have their
gradient buffer zeroed at \callulfmallreduce time (because \texttt{is\_spare} is true and \texttt{at\_boundary} is false in the steady state); they are immediately ready to absorb the next
failure, since promotion in Record simply clears
\texttt{is\_spare} and the spare's already-computed gradients were snapshotted thus can be rewound to become its contribution. The next failure that lands in a role that has a spare available therefore stays in the non-boundary branch of \callpolicyadjust, and the iteration commits without extending $P(\mathrm{major})$ at all --- exactly the
\emph{non-boundary-crossing failure} sketched in panel~(iv) of \cref{fig:workload}.

\section{Proof Sketch of the Stochastic-Equivalence Guarantee}
\label{app:proof}

We make the stochastic-equivalence property of \cref{sec:framework} precise. The claim is \emph{not} that \framework reproduces a failure-free reference run step-for-step or bitwise; rather, that as a
distribution over the same data stream, every iteration of a
\framework run draws its gradient from the same distribution as the
corresponding reference iteration. Throughout, we re-use the
notation of \cref{sec:versatile}: $W_{\mathrm{init}}, G_{\mathrm{init}},
B = W_{\mathrm{init}} \cdot G_{\mathrm{init}}$ are the initial
replica count, the initial grad-accum factor, and the target global
batch; $W_{\mathrm{cur}}(t), G_{\mathrm{cur}}(t)$ are the current
values at iteration $t$; $C_r(t)$ is the contribution count of
replica $r$ at iteration $t$ (its role-assigned microbatch count for
a major or minor, zero for a spare); and the cross-replica process
group $PG_{\mathrm{cross}}$ has world epoch
$\epsilon_{\mathrm{cur}}(t)$, incremented on every successful repair.

\paragraph{Setup.}
Fix a deterministic model with parameters $\theta_t$ at iteration
$t$ and per-example gradient $g(\theta_t; x)$, and fix a
deterministic pre-training data stream $X = (x_0, x_1, \dots)$
partitioned across the initial replicas, with each replica owning a
disjoint slice of $X$. Let $\mathcal{R}(t)$ denote the set of
replicas alive at iteration $t$, and let $\phi_t : \mathcal{R}(t)
\to \mathcal{P}(X)$ be the explicit replica-to-microbatch assignment whose gradient contributes to iteration $t$'s gradient sum assignment, with $|\phi_t(r)| =
C_r(t)$. For a major or a minor, $\phi_t(r)$ is the set of microbatches $r$ runs forward+backward on; for a major-spare or minor-spare, $\phi_t(r)=\emptyset$, since the spare's gradient is zeroed at all-reduce time. Define
\[
\mathcal{A}_{\framework}(t) \;=\; \bigsqcup_{r \in \mathcal{R}(t)} \phi_t(r),
\qquad
\mathcal{A}_{\mathrm{ref}}(t) \;=\; \text{the analogous multiset for a failure-free reference run},
\]
where $\bigsqcup$ is the disjoint union of the per-replica
assignments. Both are multisets of microbatches drawn from $X$.

\paragraph{Claim.}
Let $\Sigma$ be any failure schedule that retains at least one
replica at every $t$. Then for every iteration $t$,
$|\mathcal{A}_{\framework}(t)| = B = |\mathcal{A}_{\mathrm{ref}}(t)|$
and both multisets are subsets of $X$ without repetition. The
multisets differ only in which slice of $X$ each microbatch is
drawn from: failed replicas' slices drop out of $X$ entirely once
the replica is lost, and survivors advance through their own slices
slightly faster than the reference would prescribe. Under the
standard exchangeability assumption on the pre-training data stream,
$\mathcal{A}_{\framework}(t)$ and $\mathcal{A}_{\mathrm{ref}}(t)$
have the same distribution over $X$ for every $t$, so the iteration
gradient $\frac{1}{B} \sum_{x \in \mathcal{A}_{\framework}(t)}
g(\theta_t; x)$ is drawn from the same distribution as the
reference's $\frac{1}{B} \sum_{x \in \mathcal{A}_{\mathrm{ref}}(t)}
g(\theta_t; x)$.

\paragraph{Proof sketch.}
The protocol of \cref{sec:framework} maintains, for every iteration
$t$, the explicit replica-to-microbatch assignment $\phi_t :
\mathcal{R}(t) \to \mathcal{P}(X)$ with $|\phi_t(r)| = C_r(t)$ and
$\bigsqcup_{r} \phi_t(r) = \mathcal{A}_{\framework}(t)$. We verify
$|\mathcal{A}_{\framework}(t)| = B$ at every iteration by cases.

\emph{Case (a): no failure at $t$.} The role assignment is
inherited from iteration $t-1$. Majors and (the at most one) minor
contribute their role-assigned microbatch counts; spares contribute
$C_r(t) = 0$ but still execute forward and backward, with their
gradient buffer zeroed at \ulfmallreduce time. The world epoch is
unchanged across the iteration, so every bucket is reduced exactly
once at the same epoch $\epsilon_{\mathrm{cur}}(t)$, and
$\sum_r C_r(t) = n_{\mathrm{maj}} \cdot G_{\mathrm{cur}} +
n_{\mathrm{min}} \cdot R_{\mathrm{cur}} = B$ holds by construction
of the policy steady state.

\emph{Case (b): failure at $t$, spare available in the failed
role.} The C++ helper \texttt{record\_and\_handling\_failure}
(\cref{app:impl}) repairs $PG_{\mathrm{cross}}$, increments the
world epoch from $\epsilon_{\mathrm{cur}}$ to $\epsilon_{\mathrm{cur}}
+ 1$, and atomically promotes a spare into the vacated role, all
within the same allreduce path. The promoted spare had been zeroing
its contributions only at \ulfmallreduce time, so its gradient
bytes already exist locally; rewinding from snapshots simply admits them. Every
bucket whose snapshot was tagged with the pre-failure epoch is
rewound from its snapshot and re-reduced under
$\epsilon_{\mathrm{cur}} + 1$ via \cref{alg:gradient-restoration}, so every
bucket in the iteration is summed under the same epoch. Because
the spare's contribution count exactly replaces the failed
replica's, $\sum_{r \in \mathcal{R}(t)} C_r(t) = B$.

\emph{Case (c): failure at $t$, spares of the failed role are
exhausted (policy boundary).} The policy detects the boundary via
phase~3 of \ulfmallreduce, picks the smallest integer
$G_{\mathrm{ext}} \geq 1$ with $C_{\mathrm{cur}} + W_{\mathrm{cur}}
\cdot G_{\mathrm{ext}} \geq B$, and designates
$W_{\mathrm{cur}} \cdot G_{\mathrm{ext}} - (B - C_{\mathrm{cur}})$
survivors as boundary minors that contribute $G_{\mathrm{ext}} - 1$
extra microbatches; the other survivors contribute the full
$G_{\mathrm{ext}}$. By construction of the boundary-step counts,
\[
\sum_{r \in \mathcal{R}(t)} C_r(t)
\;=\; C_{\mathrm{cur}}
   \;+\; (\text{majors}) \cdot G_{\mathrm{ext}}
   \;+\; (\text{boundary minors}) \cdot (G_{\mathrm{ext}} - 1)
\;=\; B.
\]
As in case (b), every bucket whose snapshot tag predates the
post-repair epoch is rewound and re-reduced under
$\epsilon_{\mathrm{cur}} + 1$ via \cref{alg:gradient-restoration}, so every
bucket admitted into iteration $t$ is reduced under the same
membership and none of the failed replica's pre-failure work is
retained. The iteration gradient is therefore a uniformly-weighted
sum of exactly $B$ survivor microbatches.

The three cases together establish $|\mathcal{A}_{\framework}(t)| =
B$ for every $t$. By the disjointness of $\phi_t$, no microbatch in
$X$ is admitted twice in the same iteration. Across the run, the
multiset $\bigsqcup_t \mathcal{A}_{\framework}(t)$ is therefore a
subset of $X$ that omits the failed replicas' slices entirely;
because each replica owns a disjoint slice of $X$, dropping a
replica removes only its slice and the survivors continue to
consume their own slices in order. The pre-training stream is
effectively infinite (the corpus is never lapped), so the
remaining-survivors stream is statistically indistinguishable from a
different random shuffle of $X$. Under exchangeability of $X$,
$\mathcal{A}_{\framework}(t)$ and $\mathcal{A}_{\mathrm{ref}}(t)$
have the same distribution over $X$ at every $t$, which is the
distributional-equivalence claim.
\hfill$\square$


\section{Supplementary Evaluations}
\label{app:supp-eval}

\subsection{Evaluation under i.i.d.\ Failures}
\label{app:iid-failures}

The failure schedule of our 512-GPU experiments is not fixed: each failure is placed at a randomly chosen step within its interval, which lets the run complete all desired failures within a cluster job's wall-time. As a reference, we present an earlier 128-GPU experiment in which each device fails independently at each step with a pre-defined failure probability. The sampled failure steps are 35, 103, 118, 136, 139, 195, 252, 346 and 370; only 3 of the 9 failures fall after step 200, as fewer and fewer GPUs are alive and the sampling often hits ranks of already-failed replicas, which makes the schedule much sparser. Here each step's global batch size is only 64, much smaller than the 8192 of our 512-GPU experiment, so each step's gradient has a higher variance and each step's loss value differs slightly more. Under this i.i.d.\ failure schedule, \framework still produces a training curve statistically identical to the failure-free reference (\cref{tab:iid-loss}).

\begin{table}[h]
\centering
\small
\begin{tabular}{lccccccc}
\toprule
Step & 50 & 100 & 150 & 200 & 250 & 300 & 350 \\
\midrule
Failure-free reference & 7.66 & 6.69 & 5.98 & 5.50 & 5.31 & 4.98 & 4.75 \\
\framework & 7.65 & 6.75 & 6.02 & 5.51 & 5.24 & 4.94 & 4.74 \\
\bottomrule
\end{tabular}
\vspace{5pt}
\caption{Training loss under the i.i.d.\ failure schedule of \cref{app:iid-failures} (128 GPUs).}
\label{tab:iid-loss}
\end{table}

\subsection{Cost of In-Step Recovery}
\label{app:snapshot-cost}

Gradient snapshotting and restoration incur no additional communication, as each rank locally copies its own pre-all-reduce gradient shard. Their peak GPU-memory overhead is one local gradient copy, and their runtime is the HBM traffic divided by the memory bandwidth. For our 7B model with TP$=4$, PP$=2$, balanced partitioning and FP32 gradient accumulation, the snapshot takes $7\text{B}/(4\times2)\times4$ bytes $=3.5$\,GB per GPU; taking it reads and writes $3.5$\,GB each, i.e., about $2\times3.5/1555\approx4.5$\,ms on an A100, and the same holds for restoration. Both happen once per iteration, which lasts tens to hundreds of seconds, so their cost is negligible.


\end{document}